\documentclass[preprint]{aastex}
\usepackage{graphicx}
\usepackage{pdflscape}

\shorttitle{Legus stellar populations}
\shortauthors{Sabbi et al.}

\begin{document}
\tracingall

\title{The resolved stellar populations in the LEGUS galaxies$^1$}
\author{E. Sabbi\altaffilmark{2}, D. Calzetti\altaffilmark{3}, L. Ubeda\altaffilmark{2}, A. Adamo\altaffilmark{4},
M. Cignoni\altaffilmark{5, 6}, D. Thilker\altaffilmark{7}, A. Aloisi\altaffilmark{2}, B. G. Elmegreen\altaffilmark{8}, 
D. M. Elmegreen\altaffilmark{9}, D. A. Gouliermis\altaffilmark{10, 11}, E. K. Grebel\altaffilmark{12}, 
M. Messa\altaffilmark{4}, L. J. Smith\altaffilmark{13}, M. Tosi\altaffilmark{14}, A. Dolphin\altaffilmark{15},
J. E. Andrews\altaffilmark{16}, G. Ashworth\altaffilmark{17}, S. N. Bright\altaffilmark{2}, T. M. Brown\altaffilmark{2}, 
R. Chandar\altaffilmark{18}, C. Christian\altaffilmark{2}, G. C. Clayton\altaffilmark{19}, D. O. Cook\altaffilmark{20, 21}, 
D. A. Dale\altaffilmark{21}, S. E. de Mink\altaffilmark{22}, C. Dobbs \altaffilmark{23}, A. S. Evans \altaffilmark{24, 25}, 
M. Fumagalli \altaffilmark{17}, J. S. Gallagher III\altaffilmark{26}, K. Grasha\altaffilmark{3},  
A. Herrero\altaffilmark{27, 28}, D. A. Hunter\altaffilmark{29}, K. E. Johnson\altaffilmark{18}, L. Kahre\altaffilmark{30}, 
R. C. Kennicutt\altaffilmark{31}, H. Kim\altaffilmark{32}, M. R. Krumholz\altaffilmark{33}, J. C. Lee\altaffilmark{2}, 
D. Lennon\altaffilmark{34}, C. Martin\altaffilmark{19},  P. Nair\altaffilmark{35},  A. Nota\altaffilmark{13}, 
G. \"Ostlin\altaffilmark{7}, A. Pellerin\altaffilmark{36}, J. Prieto\altaffilmark{37}, M. W. Regan\altaffilmark{2},  
J. E. Ryon\altaffilmark{2}, E. Sacchi\altaffilmark{14,38}, D. Schaerer\altaffilmark{39}, D. Schiminovich\altaffilmark{40}, 
F. Shabani\altaffilmark{12}, S. D. Van Dyk\altaffilmark{41}, 
R. Walterbos\altaffilmark{30}, B. C. Whitmore\altaffilmark{2}, A. Wofford\altaffilmark{42},  
}
\email{sabbi@stsci.edu}

\altaffiltext{1}{Based on observations with the NASA/ESA Hubble Space Telescope, obtained at the Space Telescope Science Institute, which is operated by AURA Inc., under NASA contract NAS 5-26555}
\altaffiltext{2}{Space Telescope Science Institute, Baltimore, MD, USA}
\altaffiltext{3}{Department of Astronomy, University of Massachusetts, Amherst, MA 01003, USA}
\altaffiltext{4}{Department of Astronomy, Oskar Klein Centre, Stockholm University, AlbaNova University Centre, SE-106 91 Stockholm, Sweden}
\altaffiltext{5}{Department of Physics, University of Pisa, Largo B. Pontecorvo 3, I-56127, Pisa, Italy}
\altaffiltext{6}{INFN, Largo B. Pontecorvo 3, I-56127, Pisa, Italy}
\altaffiltext{7}{Department of Physics and Astronomy, The Johns Hopkins University, Baltimore, MD, USA}
\altaffiltext{8}{IBM Research Division, T.J. Watson Research Center, Yorktown Heights, NY, USA}
\altaffiltext{9}{Department of Physics and Astronomy, Vassar College, Poughkeepsie, NY, USA}
\altaffiltext{10}{Zentrum f\"ür Astronomie der Universit\"ät Heidelberg, Institut f\"ür Theoretische Astrophysik, Albert-Ueberle-Str. 2, D-69120 Heidelberg, Germany}
\altaffiltext{11}{Max Planck Institute for Astronomy, K\"önigstuhl 17, D-69117 Heidelberg, Germany}
\altaffiltext{12}{Astronomisches Rechen-Institut, Zentrum f\"ür Astronomie der Universit\"ät Heidelberg, M\"önchhofstr. 12-14, D-69120 Heidelberg, Germany}
\altaffiltext{13}{European Space Agency/Space Telescope Science Institute, Baltimore, MD, USA}
\altaffiltext{14}{INAF--Osservatorio Astronomico di Bologna, Bologna, Italy}
\altaffiltext{15}{Raytheon Company, 1151 East Hermans Road, Tucson, AZ 85756, USA}
\altaffiltext{16}{Department of Astronomy, University of Arizona, Tucson, AZ, USA}
\altaffiltext{17}{Institute for Computational Cosmology and Centre for Extragalactic Astronomy, Department of Physics, Durham University, Durham, UK}
\altaffiltext{18}{Department of Physics and Astronomy, University of Toledo, Toledo, OH, USA}
\altaffiltext{19}{Department of Physics and Astronomy, Louisiana State University, Baton Rouge, LA, USA}
\altaffiltext{20}{California Institute of Technology, Pasadena, CA, USA}
\altaffiltext{21}{Department of Physics and Astronomy, University of Wyoming, Laramie, WY, USA}
\altaffiltext{22}{Astronomical Institute Anton Pannekoek, Amsterdam University, Amsterdam, The Netherlands}
\altaffiltext{23}{School of Physics and Astronomy, University of Exeter, Exeter, UK}
\altaffiltext{24}{Department of Astronomy, University of Virginia, Charlottesville, VA, USA}
\altaffiltext{25}{National Radio Astronomy Observatory, Charlottesville, VA, USA}
\altaffiltext{26}{Department of Astronomy, University of Wisconsin–Madison, Madison, WI, USA}
\altaffiltext{27}{Instituto de Astrofisica de Canarias, La Laguna, Tenerife, Spain}
\altaffiltext{28}{Departamento de Astrofisica, Universidad de La Laguna, Tenerife, Spain}
\altaffiltext{29}{Lowell Observatory, Flagstaff, AZ, USA}
\altaffiltext{30}{Department of Astronomy, New Mexico State University, Las Cruces, NM, USA}
\altaffiltext{31}{Institute of Astronomy, University of Cambridge, Cambridge, UK}
\altaffiltext{32}{Gemini Observatory, La Serena, Chile}
\altaffiltext{33}{Research School of Astronomy and Astrophysics, Australian National University, Canberra, ACT Australia}
\altaffiltext{34}{European Space Astronomy Centre, ESA, Villanueva de la Ca\~nada, Madrid, Spain}
\altaffiltext{35}{Department of Physics and Astronomy, University of Alabama, Tuscaloosa, AL, USA}
\altaffiltext{36}{Department of Physics and Astronomy, State University of New York at Geneseo, Geneseo, NY, USA}
\altaffiltext{37}{Department of Astrophysical Sciences, Princeton University, Princeton, NJ, USA}
\altaffiltext{38}{Department of Physics and Astronomy, Bologna University, Bologna, Italy}
\altaffiltext{39}{Observatoire de Genève, University of Geneva, Geneva, Switzerland}
\altaffiltext{40}{Department of Astronomy, Columbia University, New York, NY, USA}
\altaffiltext{41}{IPAC/CalTech, Pasadena, CA, USA}
\altaffiltext{42}{Instituto de Astronom\'ia, Universidad Nacional Aut\'onoma de M\'exico, Unidad Acad́\'emica en Ensenada, Km 103 Carr. Tijuana-Ensenada, Ensenada 22860, M\'exico }

\begin{abstract}
The Legacy ExtraGalactic UV Survey (LEGUS) is a multiwavelength Cycle 21 Treasury program on the Hubble Space Telescope. It 
studied 50 nearby star-forming galaxies in five bands from the near UV to the I-band, combining new Wide Field Camera 3 
observations with archival Advanced Camera for Surveys data. LEGUS was designed to investigate how star formation occurs and 
develops on both small and large scales, and how it relates to the galactic environments. In this paper we present the 
photometric catalogs for all the apparently single stars identified in the 50 LEGUS galaxies. Photometric catalogs 
and mosaicked images for all filters are available for download. 

We present optical and near UV color-magnitude diagrams for all the galaxies. For each galaxy we derived the distance from 
the tip of the red giant branch. We then used the NUV color-magnitude diagrams to identify stars more massive than 
14 M$_\odot$, and compared their number with the number of massive stars expected from the GALEX FUV luminosity. Our analysis 
shows that the fraction of massive stars forming in star clusters and stellar associations is about constant with the star formation rate. 
This lack of a relation suggests that the time scale for evaporation of unbound structures is comparable or longer than 10 Myr.
At low star formation rates this translates to an excess of mass in clustered environments as compared to model predictions 
of cluster evolution, suggesting that a significant fraction of stars form in unbound systems. 
\end{abstract}


\section{Introduction}

Star formation (SF) plays a major role in shaping the evolution of galaxies. On a small scale, feedback from massive stars 
affects the surrounding environment via intense stellar winds, ultraviolet radiation fields, chemical processing, and explosions. 
On a large scale, it governs the macroscopic properties of galaxies. Despite its importance, how galaxies form stars over time, 
likely in response to both internal and external factors, has not been characterized, nor have we fully understood the link 
between SF and the global properties of the host galaxies. As a result a universal law that describes SF at all scales is still 
missing \citep{dobbs11, dobbs13, hopkins13}.

The fragmentation of giant molecular clouds (GMCs) likely causes the formation of stars in a clustered environment 
\citep[e.g.,][]{larson81, elmegreen04, maclow04, mckee07}, and it has been known for a while that a fraction of the newly formed 
systems is not gravitationally bound \citep{blaauw64, elmegreen83, clarke00, lada03, portegies10, goddard10, gieles11}. For a 
long time it was believed that the sudden expulsion of residual gas by feedback was decimating the population of young embedded 
stellar systems \citep[e.g., infant mortality,][]{tutkov78, hills80, lada03, bastian06, parmentier08}, but the strong 
correlation between young stellar objects (YSOs) and the hierarchical structure of the interstellar medium 
\citep[ISM,][]{testi00, gutermuth11, bressert12} suggests that only in a fraction of cases the star formation efficiency is 
sufficiently high to result in bound stellar systems, with the majority of the stars forming in more dispersed structures 
throughout the natal GMC \citep{elmegreen01}. This suggests that, at least in some environments, SF may occur on a continuous 
spectrum of number densities, without the need for a critical density threshold \citep[e.g.][]{bressert10, parker12} . 

The fraction of stars that form in bound star clusters is often quantified as the cluster formation efficiency $\Gamma$ 
\citep[e.g.,][]{bastian08, goddard10, adamo11, silvavilla11}. This parameter provides important information on the process of 
SF \citep{elmegreen02} and cluster disruption \citep{gieles05} in different environments, and it is also a powerful tracer of 
the history of star formation in distant galaxies \citep{miller97, larsen01, goudfrooij04, bastian05, smith07, fedotov11}.

Observational studies of extragalactic cluster populations highlighted a correlation between $\Gamma$ and the star formation 
rate surface density  $\Sigma_{SFR}$ of the host galaxy \citep{larsen00, goddard10, adamo11,silvavilla11}. This correlation 
seems to suggest that the short free-fall times characteristic of the regions of higher gas density would be sufficient to 
enable the high star formation efficiencies necessary to form bound systems \citep{kruijseen12}.

Several studies have investigated if and how $\Gamma$ depends on the environment. While in investigating how $\Gamma$ changes 
within a galaxy, the sample of clusters used to derive $\Gamma$ was often carefully characterized \citep[e.g.][]{adamo15, johnson16, 
adamo17, messa18}, when comparing results from different galaxies, this has often resulted in combining heterogeneous samples of clusters, 
whose ages and properties had been derived in non-uniform ways. In this paper we take advantage of the Legacy ExtraGalactic UV 
Survey (LEGUS, PI Calzetti, GO-13364) to investigate the relation between young massive stars (age $<14$ Myr) in the field of 
the host galaxy and the corresponding population of young star clusters and associations. 

LEGUS is a Cycle 21 Treasury program that images 50 nearby (distance $\sim 3.5-18$ Mpc) star-forming galaxies in five bands (NUV, U, B, V, I) with the Hubble Space Telescope (HST). 
LEGUS targets were carefully chosen to span the widest range of morphology, star formation rate (SFR), mass, metallicity, 
internal structure, and interaction state. For each of the targets ancillary data from the near- ($\lambda=0.231 \mu$m) and 
far-UV ($\lambda=0.153 \mu$m) from {\it GALEX}, ground based H$\alpha$+[N{\sc ii}], and from Spitzer Space Telescope and WISE 
mosaics ($\lambda \lambda=$3.4 -- 160$\mu$m) are available \citep[a complete description of the LEGUS sample and goals is 
presented in][]{calzetti15}. 

The paper is organized as follows: Section~\ref{The Data} describes the observations and the data reduction; the color-magnitude 
diagrams (CMDs) are presented in Section~\ref{CMDs}, and the distances of our targets as derived from the tip of the red giant 
branch (TRGB) are discussed in Section~\ref{trgb}. Section~\ref{gamma} discusses $\Gamma$ as derived from the fraction of massive 
stars in the field ($\Gamma^\star$) as a function of the SFR estimated from the far UV emission. The summary and conclusions are 
presented in Section~\ref{conclusions}.

\section{Observations and Data Reduction}
\label{The Data}
\subsection{The Data}

LEGUS was awarded 154 orbits in Cycle 21 to observe 50 star-forming galaxies with the UVIS channel of the Wide Field Camera 3 
(WFC3) in the filters F275W, F336W, and, when not already available in the Mikulski Archive for the Space Telescope (MAST) 
archive, also in the filters F438W, F555W, and F814W. For simplicity, from now on, in the text we will refer to these filters 
as NUV, U, B, V, and I respectively. The choice of filters was dictated by the desire to distinguish young massive bright stars 
from faint star clusters, derive accurate star formation histories for the stars in the field from CMDs, and obtain 
extinction-free ages and masses for the star clusters.   

Nearly half of the LEGUS galaxies are sufficiently compact that their far-UV (FUV) effective surface can be mapped with one 
WFC3/UVIS pointing. For those cases in which the galaxy is slightly larger than the UVIS field of view (FoV), the pointing was 
optimized to achieve the best coverage possible. Similarly, we tuned the observation position angle to maximize the overlap with 
archival data (when available). Eleven of the LEGUS galaxies are significantly more extended than the WFC3 FoV. For nine of 
them we collected multiple pointings along the radial direction to span a wider range of environments. 

For each filter we acquired three observations following a three-point dither-pattern with a step of $\sim 1\farcs88$. We 
devoted  a total integration time of $\ge 2400s$ in the NUV filter and $\ge 1100s$ in U. For 15 targets, 
where long B, V, and I ACS observations were already available in the MAST archive, the final exposure time in U went up to 
$\ge2400s$. For those galaxies that were not previously observed with ACS in the optical, we used $\ge 900 s$ to observe them 
in B and I, and $\ge 1100s$ to acquire data in V. The observations were designed to reach a depth of $m_{Vega}$(NUV)$=24.5$ 
with signal-to-noise ratio (S/N) $\sim 6$ and comparable depth in the other filters. The log of the observations, complete with 
dataset names, proposal IDs, observing dates and times, coordinates, instruments, filters, and exposure times, is reported in 
Appendix A/Table~\ref{log}.   

After five years of operations, charge transfer efficiency (CTE) losses were becoming a concern \citep{bourque13} for WFC3/UVIS,
especially at the shorter wavelengths, where the background is low \citep{baggett12}. To mitigate the effect of the degrading 
CTE, we artificially increased the background level of the NUV, U, and B exposures by 12 e$^-$ using post-flash. Bias level, 
superbias, superdark subtraction, post-flash removal, and flat-field correction were applied to each WFC3 image through the 
standard calibration pipeline CALWF3 version 3.1.2. At the time of the data processing, the pixel-based CTE correction was 
not included in CALWF3, therefore we used the standalone software routine {\tt wfc3uv\_ctereverse\_parallel.F} 
provided by the WFC3 team (available at the webpage http://www.stsci.edu/hst/wfc3/tools/cte\_tools). 
Pipeline-processed and CTE-corrected ACS/WFC $*\_${\tt flc}.fits individual images were downloaded directly from the MAST 
archive.

\subsection{Image alignment}
\label{drizzle}

On average the World Coordinate System (WCS) solution in the header of WFC3 data is more accurate than the one found in old ACS 
datasets. Therefore we decided to use the information found in the WFC3 B-band images to align and register the individual 
$*\_${\tt flc} images with north up and east to the left. When WFC3 B-band data were not available we used the U-band filter 
as our reference frame. 

Each image was aligned using the {\tt tweakreg} routine. The {\tt astrodrizzle} routine \citep{gonzaga12} was then used to 
combine the aligned images for each filter. Each final drizzled image is sky-subtracted and weighted by its exposure time; it 
is in units of $e^- s^{-1}$ and has a pixel scale of 39.62 mas pixel$^{-1}$. The routine {\tt tweakback} was then used to 
propagate the improved World Coordinate System solution back to the header of the $*\_${\tt flc} images. Astrodrizzle was also 
run to identify those pixels that were affected by cosmic rays (CRs) and other spurious signals (such as hot or bad pixels). 
This information was blotted back to the data quality ({\tt DQ}) arrays of the affected original frames.

The ACS/WFC data were aligned and drizzled to the UVIS pixel scale. Because of different orientations and pointings, in some 
cases we had to mosaic multiple ACS images together to maximize the overlap with the WFC3 data. All images are available for 
download at 

https://archive.stsci.edu/prepds/legus/dataproducts-public.html

\subsection{Photometric Reduction}

Positions and fluxes for point-like sources were measured via PSF-fitting using the WFC3 and ACS modules of the photometry 
package DOLPHOT version 2.0, downloaded on 12 December 2014 from the website http://americano.dolphinsim.com/dolphot/ 
\citep{dolphin02}. Each filter was analyzed independently of the others to both maximize the sensitivity to objects of extreme 
colors, and expedite the data reduction. 

The fluxes of all the sources exceeding an initial sigma detection threshold {\tt SigFind}$\ga 3.0$ in the drizzled images were 
measured on the individual $*\_${\tt flc} frames. The final photometric analysis was then performed on all the sources with 
sigma threshold {\tt SigFinal}$\ga 3.5$ . 

For the WFC3 data we used the set of PSF libraries whose cores are based on the effective PSF published by the WFC3 team 
(see the WFC3 website for further information\footnote{http://www.stsci.edu/hst/wfc3/analysis/PSF}), while for the ACS data we 
used the PSF libraries derived from Tiny Tim \citep{krist95}. For each filter the astrometric solution stored in the header of 
the $*\_${\tt flc} was used as the starting point for the alignment of the images in a filter, and then we allowed DOLPHOT to 
use the bright stars in common to the images to further refine the solution. The final alignment among each image was good to 
$\sim 0\farcs01$ precision. The same stars were also used to improve the reference PSFs, by minimizing the residuals from the 
bright stars fitting. The magnitude of each detected star was iteratively improved by taking into account the contribution of 
the nearby sources.

We ran a series of artificial star tests on a subset of galaxies, characterized by different amounts of crowding and background 
levels to optimize the DOLPHOT setup. The final set of parameters we used in our analysis is listed in Table~\ref{dol_par}. 
Note that because we had applied the pixel-based CTE correction to each individual $*\_${\tt flt} image, we turned off the 
DOLPHOT routine that empirically corrects the photometry for the CTE losses. 

\begin{deluxetable}{rcl}
\tablecolumns{3}
\tabletypesize{\scriptsize}
\tablewidth{0pt}
\tablecaption{Parameters used to run the stellar photometry with DOLPHOT\label{dol_par}} 
\startdata
\hline\hline
\multicolumn{3}{}{}\\
\multicolumn{1}{}{}& \multicolumn{1}{c}{DOLPHOT PARAMETERS} & \multicolumn{1}{}{} \\
\multicolumn{3}{}{}\\
\hline
img?\_shift &=& 0 0 \\ 
img?\_xform &=& 1 0 0 \\
img?\_apsky &=& 15 25 \\
img?\_RAper &=& 8 \\
img?\_RPSF &=& 10 \\
img?\_RSky &=& 15.0 35.0\\ 
img?\_RChi &=& 2.0 \\
RSky0 &=& 15\\
RSky1 &=& 35\\
RPSF &=& 10\\
RCentroid &=& 2\\
SigFind &=& 3.0\\
SigFindMult &=& 0.85\\ 
SigFinal &=& 3.5 \\
MaxIT &=& 25 \\
PSFPhot &=& 1 \\
FitSky &=& 3 \\
SkipSky &=& 2 \\
SkySig &=& 2.25 \\
NoiseMult &=& 0.10\\
FSat &=& 0.999\\
PosStep &=& 0.25\\
dPosMax &=& 2.5\\
RCombine &=& 1.5\\
SigPSF &=& 5.0\\
UseWCS &=& 1\\
Align &=& 3\\
WFC3useCTE &=& 0\\
AlignIter &=& 2\\
Rotate &=& 1\\
SecondPass &=& 1\\
Force1 &=& 1\\
PSFres &=& 1\\
psfoff &=& 0.0\\
ApCor &=& 1\\
WFC3UVISpsfType &=& 1 \\
\enddata
\end{deluxetable}

We recommend users interested in reproducing LEGUS photometry to download both the drizzled-combined $*\_${\tt drc} and 
individual CTE-corrected $*\_${\tt flc} images from the LEGUS webpage 
https://archive.stsci.edu/prepds/legus/dataproducts-public.html because DOLPHOT uses the information stored in the {\tt DQ} 
arrays of the $*\_${\tt flc} images to discard bad pixels and cosmic ray hits. This information is not recorded in the data 
downloaded from the MAST archive, and this difference can cause up to a $5\%$ difference in the final photometry. In addition, 
the WFC3 team has recently largely modified the calibration pipeline, to better reflect the difference in quantum efficiency 
between the two chips of the UVIS channel \citep{Ryan16}. Our analysis was performed before the changes in the pipeline were 
introduced.

DOLPHOT went through several changes and updates in the time we processed our dataset. These changes did not always correspond 
to a change in the package version number. To guarantee self-consistent results for users interested in reproducing LEGUS 
results, or in running tailored artificial star tests, the DOLPHOT version used to reprocess the entire LEGUS dataset, as well as 
the version of PSF libraries can be downloaded from the LEGUS webpage. The python-based pipeline to automate the photometry is 
also available for download.
 
\subsection{The photometric catalogs}

For each star DOLPHOT provides the position relative to the drizzled image, the magnitude and a series of diagnostics to 
evaluate the quality of the photometry, including S/N ratio, photometric error, $\chi^2$ for the fit of the PSF, roundness 
(which can be used to identify extended objects), object type (which describes the shape of the source), and an error flag, 
which is larger than zero whenever there is an issue with the fitting (i.e., because of saturation or extension of the source 
beyond the detector field of view).

For each galaxy we are releasing several different photometric catalogs that can be downloaded from the MAST archive. 
In particular we are releasing:
\begin{itemize}
\item The five (one for each filter) single-band outputs of DOLPHOT. 
\item A five band catalog that includes all the sources with {\tt ERRORFLAG}$=$0, 1, 2 or 3, and magnitude brighter than 28.5 and has been
labeled as v1;
\item A five band catalog that comprises only sources with magnitude brighter than 30 and with {\tt ERRORFLAG}$=$0. This catalog has 
been labeled as v2.
\end{itemize}



In both versions v1 and  v2, the five single-band catalogs were cross-matched using the cross-correlation 
package CataPack\footnote{The CataPack package is developed by P. Montegriffo at the Bologna  Observatory (INAF), and is 
available for download at http://www.bo.astro.it/$\sim$paolo/Main/CataPack.html}. Because of the wide range of wavelengths 
covered by LEGUS, we used the intermediate B filter as the reference frame. 

\subsection{Selection criteria}

\begin{figure}
\epsscale{1.0}
\plotone{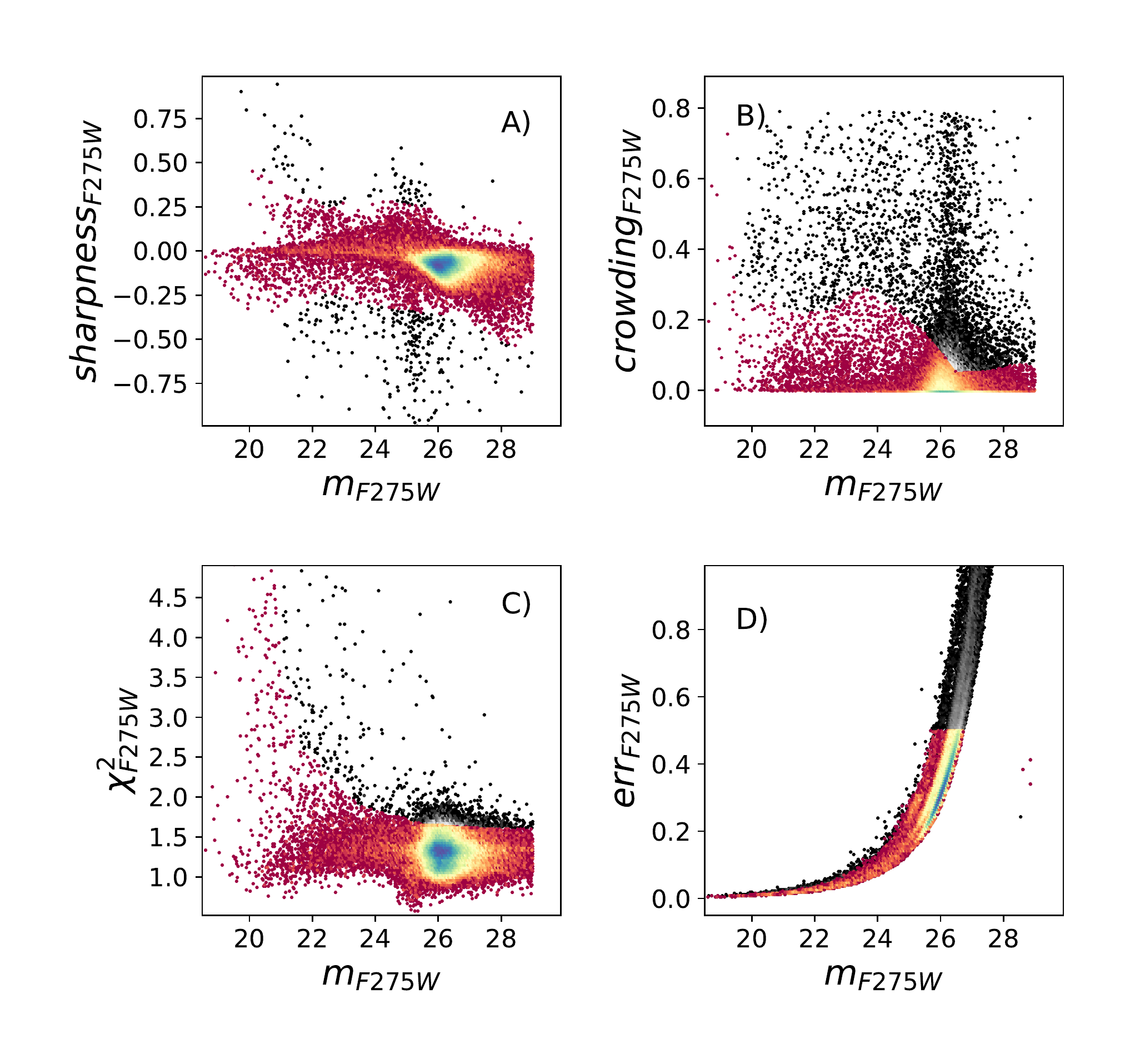}
\caption{\label{f:sel} Quality of the photometry for the sources detected by DOLPHOT in the NUV filter for the dwarf galaxy 
NGC~1705. In each plot all the detected sources are shown as light gray dots, while the two-dimensional histograms highlight 
the distribution of the sources that passed the selection criteria. Sources were first selected on the basis of their 
sharpness - {\it Panel A)}, the local crowding - {\it Panel B)}, then their $\chi^2$ value - {Panel C)}, and roundness - {\it Panel D)}.}
\end{figure}

To correctly interpret the properties of the stellar populations found in the LEGUS galaxies, we need to differentiate as much 
as possible the bona fide apparently single stars from extended objects, blended sources, and spurious detections. To achieve 
this goal we independently applied to each filter selection criteria based on the photometric error, the shape of the objects,
their isolation and the quality of the PSF-fitting, and the roundness. 

In both versions of the multiband photometric catalogs we have used a five digit (one digit per filter from NUV to I from left to right) 
good photometric quality flag ({\tt GPQF}). In version v1 each digit can range from 0 to 7, and in version v2 it can range from 0 to 5. In both
catalogs 0 correspondis to a non-detection. From the sources with magnitude brighter than 30 we then considered only those whose sharpness deviates 
less than $1\sigma$ from the mean value per magnitude bin. The sources that passed this criterion were flagged with {\tt GPQF}$ge 2$, while {\tt GPQF}$=1$
means that either the sharpness deviates more than $1\sigma$ from the mean value for that bean of magnitude. {\it Panel A)} of Figure~\ref{f:sel} 
shows how the sharpness changes as a function of magnitude in the NUV filter for the galaxy NGC~1705. The black points indicate all the sources 
that are included in the version v1 of the photometric catalog, while the two-dimensional histogram shows the sources that passed the selection in 
sharpness. 

From all the sources with ``good'' sharpness we then selected those whose crowding parameter is less then $2\sigma$ from 
the mean value per magnitude bin, and assigned them {\tt GPQF}$ge 3$. The 2D histogram in {\it Panel B)} shows the sources that passed 
both the selection in sharpness and crowding, superimposed on the distribution of the all the sources.   
 
Among the sources that passed the selection with crowding, those that are within $3\sigma$ from the mean $\chi^2$ value in the magnitude bin, 
have been flagged with {\tt GPQF}$\ge 4$ (see {\it Panel C)}), and among these we assigned  {\tt GPQF}$=5$ to the sources with photometric error
$< 0.5$ (see {\it Panel D)}) A source that passed all the selection criteria in all five bands therefore will 
have {\tt GPQF}$=55555$, while a source that was detected only in V and I, and that in both filters passed all the selection criteria 
up to the $\chi^2$ will have {\tt GPQF}$=00044$. 

In the version v1 of the catalog, we applied two further selections: sources with {\tt GPQF}$\ge 6$ are within $1\sigma$ from the mean roundness, 
and source if their S/N ration is $>2.5$ they have been flagged with {\tt GPQF}$= 7$. For each target the version v1 photometric catalog has the 
following format: x and y coordinates in the drizzled image reference frame. For each filter we report the magnitude measured in the 
VEGA system, the photometric error, the reduced $\chi^2$ from the PSF-fitting, sharpness, crowding, roundness, signal to noise ration and 
object type. Filters are listed in order of increasing wavelengths from the NUV to the I band. The {\tt ERRORFLAG} determined by DOLPHOT have been 
summarized in column 43 by a 5 digit (one for each filter) ERROR\_FLAG. Because 0 corresponds to a non detection in that filter, we added 1 to
the value of the DOLPHOT {\tt ERRORFLAG}. {\tt GPQF} is listed in column 44, while right ascension and declination are in columns 
45 and 46 respectively. The last column lists a LEGUS-unique identifier. The header of a v1 catalog is shown in Table~\ref{t:header}. 
The version v2 catalogs have the same format, expect for the flux, that is not reported.

\begin{deluxetable}{lrlll}
\tablecolumns{5}
\tabletypesize{\tiny}
\tablecaption{Header of one of the LEGUS photometric catalogs\label{t:header}} 
\startdata
\hline\hline
\# &  1 &  X\_IMAGE      &  Object position along x                           &  [pixel] \\	
\# &  2 &  Y\_IMAGE      &  Object position along y                           &  [pixel]	\\
\# &  3 &  MAGNI\_F275W  &  WFC3/UVIS/F275W DOLPHOT magnitude in VEGA system  &  [mag]	\\
\# &  4 &  MAGER\_F275W  &  Magnitude error                                   &  [mag]	\\
\# &  5 &  CHISQ\_F275W  &  Reduced chi2 from PSF-fitting                     &  []	\\
\# &  6 &  SHARP\_F275W  &  Object sharpness                                  &  []	\\
\# &  7 &  CROWD\_F275W  &  Object crowding                                   &  [mag]\\	
\# &  8 &  ROUND\_F275W  &  Object roundness                                  &  []	\\
\# &  9 &  SN\_F275W     &  Object signal to noise ratio                      &  []	\\
\# & 10 &  OTYPE\_F275W  &  Object type                                       &  []	\\
\# & 11 &  MAGNI\_F336W  &  WFC3/UVIS/F336W DOLPHOT magnitude in VEGA system  &  [mag]\\	
\# & 12 &  MAGER\_F336W  &  Magnitude error                                   &  [mag]\\	
\# & 13 &  CHISQ\_F336W  &  Reduced chi2 from PSF-fitting                     &  []	\\
\# & 14 &  SHARP\_F336W  &  Object sharpness                                  &  []	\\
\# & 15 &  CROWD\_F336W  &  Object crowding                                   &  [mag]\\	
\# & 16 &  ROUND\_F336W  &  Object roundness                                  &  []	\\
\# & 17 &  SN\_F336W     &  Object signal to noise ratio                      &  []	\\
\# & 18 &  OTYPE\_F336W  &  Object type                                       &  []	\\
\# & 19 &  MAGNI\_F438W  &  WFC3/UVIS/F438W DOLPHOT magnitude in VEGA system  &  [mag]\\	
\# & 20 &  MAGER\_F438W  &  Magnitude error                                   &  [mag]\\	
\# & 21 &  CHISQ\_F438W  &  Reduced chi2 from PSF-fitting                     &  []	\\
\# & 22 &  SHARP\_F438W  &  Object sharpness                                  &  []	\\
\# & 23 &  CROWD\_F438W  &  Object crowding                                   &  [mag]\\	
\# & 24 &  ROUND\_F438W  &  Object roundness                                  &  []	\\
\# & 25 &  SN\_F438W     &  Object signal to noise ratio                      &  []	\\
\# & 26 &  OTYPE\_F438W  &  Object type                                       &  []	\\
\# & 27 &  MAGNI\_F555W  &  WFC3/UVIS/F555W DOLPHOT magnitude in VEGA system  &  [mag]\\	
\# & 28 &  MAGER\_F555W  &  Magnitude error                                   &  [mag]\\	
\# & 29 &  CHISQ\_F555W  &  Reduced chi2 from PSF-fitting                     &  []	\\
\# & 30 &  SHARP\_F555W  &  Object sharpness                                  &  []	\\
\# & 31 &  CROWD\_F555W  &  Object crowding                                   &  [mag]\\	
\# & 32 &  ROUND\_F555W  &  Object roundness                                  &  []	\\
\# & 33 &  SN\_F555W     &  Object signal to noise ratio                      &  []	\\
\# & 34 &  OTYPE\_F555W  &  Object type                                       &  []	\\
\# & 35 &  MAGNI\_F814W  &  WFC3/UVIS/F814W DOLPHOT magnitude in VEGA system  &  [mag]\\	
\# & 36 &  MAGER\_F814W  &  Magnitude error                                   &  [mag]\\	
\# & 37 &  CHISQ\_F814W  &  Reduced chi2 from PSF-fitting                     &  []	\\
\# & 38 &  SHARP\_F814W  &  Object sharpness                                  &  []	\\
\# & 39 &  CROWD\_F814W  &  Object crowding                                   &  [mag]\\	
\# & 40 &  ROUND\_F814W  &  Object roundness                                  &  []	\\
\# & 41 &  SN\_F814W     &  Object signal to noise ratio                      &  []	\\
\# & 42 &  OTYPE\_F814W  &  Object type                                       &  []	\\
\# & 43 &  ERROR\_FLAG   &  DOLPHOT flags                                     &  [0-4] \\ 
\# & 44 &  GPQF          &  Good photometric quality flags                    &  [0-5] \\ 
\# & 45 &  ALPHA\_J2000  &  Right ascension of barycenter (J2000)             &  [deg]\\	
\# & 46 &  DELTA\_J2000  &  Declination of barycenter (J2000)                 &  [deg]\\	
\# & 47 &  LEGUS\_ID     &  LEGUS unique identifier                           &  [LEGUS+HH:MM:SS.SS-DD:MM:SS.S]	\\
\enddata
\end{deluxetable}

\section{Color-Magnitude Diagrams}
\label{CMDs}

\begin{figure}
\epsscale{1.0}
\plotone{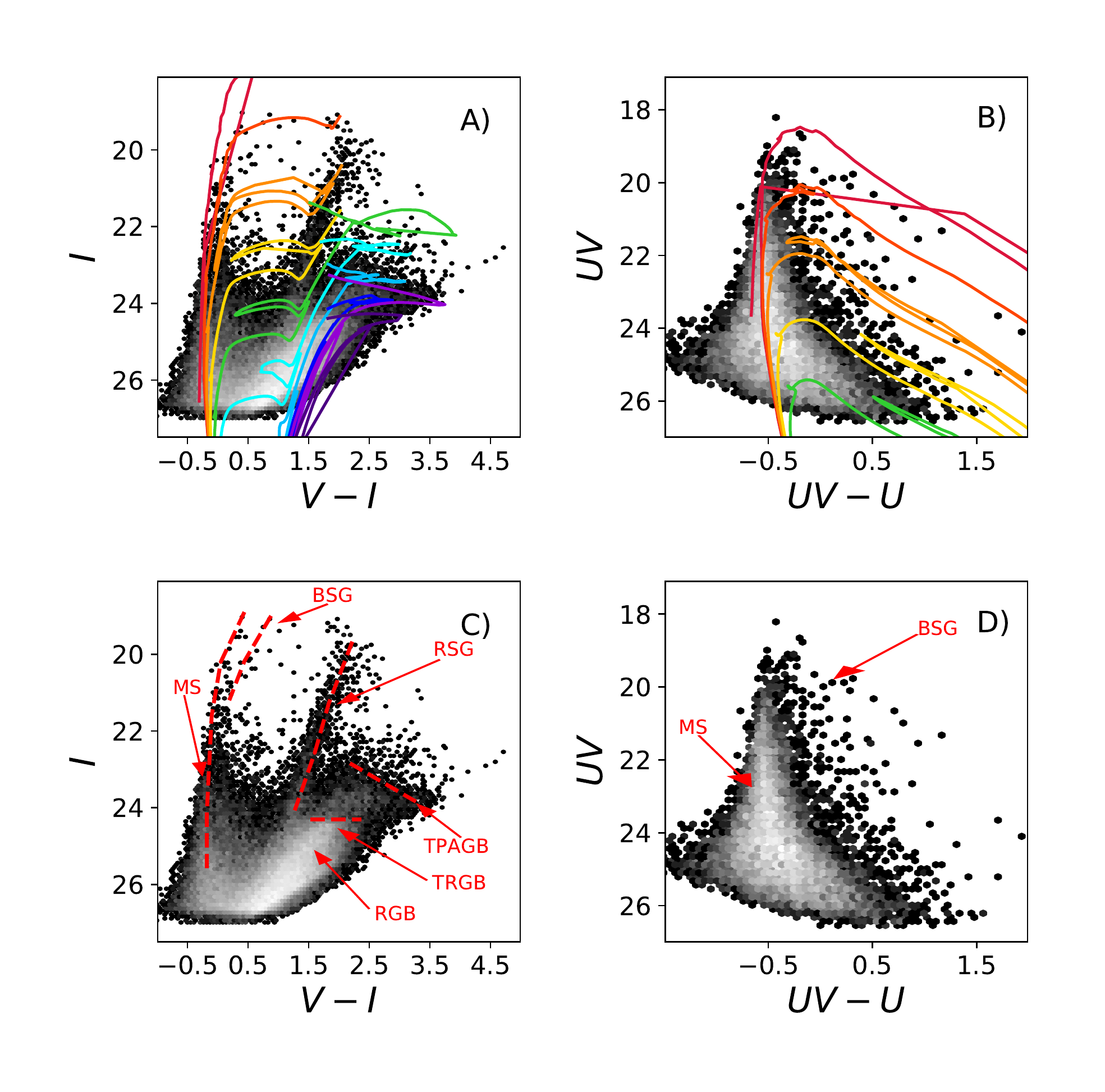}
\caption{\label{f:cmd.4395} {\it Panel A):} Optical I vs. V-I CMD for the galaxy NGC~4395. PARSEC isochrones for 4, 8, 16, 40, 100, 
250, 630 Myr and 1.6, 4, and 10 Gyr are superimposed in red, dark orange, orange, yellow, green, cyan, light blue, dark blue, 
purple and indigo respectively, by assuming a distance modulus of 28.36, and an extinction E(B-V)=0.04.  
{\it Panel B):} NUV vs NUV-U CMD for NGC~4395. PARSEC isochrones for 4, 8, 16, 
40, 100 Myr are shown in red, dark orange, orange, yellow, and green. {\it Panel C):} Optical I vs. V-I CMD with the names and position 
of the main evolutionary sequences highlighted by red-dashed lines. {\it Panel D):} NUV vs NUV-U CMD. Only the upper part of the MS and 
BSG stars are visible. Their locations are roughly indicated by red arrows.}
\end{figure}

CMDs are very effective tools to investigate the stellar content of a system, and therefore to infer its 
star formation history and evolution. Figure~\ref{f:cmd.4395}  shows the optical (I vs V-I, {\it to the left}) and NUV (NUV vs NUV-U, 
{\it to the right}) CMDs of the galaxy NGC~4605, obtained from the catalog 
{\tt hlsp\_legus\_hst\_uvis\_ngc4605\_multiband\_v2\_stellar\_phot.cat}. The left panels show only stars with {\tt GPQF}$=***55$,
while the right panels show only the sources with {\tt GPQF}$=55***$.

PARSEC isochrones \citep{bressan12, tang14, chen15, marigo17} of different ages are superimposed on the CMDs in the two upper panels. 
In the two lower panels the evolutionary features characteristic of the two CMDs are highlighted for guiding the interpretation. 

In the optical CMD stars brighter than $I<23.5$ are distributed along two sequences. The bluer sequence ($V-I<0.5$ , sometimes 
called ``{\it blue plume}'') hosts intermediate and high-mass young (down to few Myr) main-sequence (MS) stars, and core 
helium-burning stars at the blue edge of the blue loop (blue supergiants, BSGs). The bright red stars ($V-I>1.5$, {\it red plume}) 
are evolved helium-burning red supergiants (RSGs). Intermediate-mass asymptotic giant branch (AGB) stars of a few hundred million years 
can be seen in all the LEGUS optical CMDs below the red plume. In most of the CMDs it is also possible to note a sharp 
discontinuity in the luminosity function of the red stars. This corresponds to the tip of the old (age $\ga 1$ Gyr) red giant 
branch (TRGB). 

The morphology and difference in color between the red and the blue plumes, as well as the shape of RGB and tip of the 
asymptotic giant branch (TPAGB), can be used to constrain the metallicity of a galaxy. By comparing the optical CMDs to PARSEC 
isochrones, we were able to divide the galaxies in the LEGUS sample into six groups: very metal-poor galaxies with metallicity 
($Z\simeq 0.0008$), galaxies with metallicity close the Small Magellanic Cloud ($Z\simeq 0.004$), galaxies close to the 
metallicity of the Large Magellanic Cloud (LMC, $Z\simeq 0.008$), systems whose metallicity is clearly higher that in the LMC, 
but not yet solar ($Z\simeq 0.01$), galaxies with solar metallicity ($Z\simeq 0.02$), and finally super-solar systems 
($Z\ga 0.03$). In 78\% of the cases our estimates of the metallicity are in good agreement with literature values \citep[see 
Table~1 in][]{calzetti15}. In 5 cases (namely NGC~1313, NGC~1705, NGC~3344, NGC~4248, and NGC~6744) our estimates are significantly higher 
than what would be derived from the oxygen abundance reported in the literature, while in three cases (NGC~2500, NGC~4605, and NGC~5238) 
the morphology of the TRGB and the separation between the blue and the red plumes appear to be too tight compared to the oxygen 
 abundances reported in the literature.   

The NUV CMD contains high-mass MS stars and evolved blue supergiants (BSG) at the blue edge of the blue loop. The NUV and U 
filters are very effective to highlight the sites of the most recent star formation across the various galaxies. On the other 
hand at these wavelengths, even for the nearest galaxies, the LEGUS detection threshold limits the lookback time to $\le 100$ Myr.  

NUV and optical CMDs for all the LEGUS galaxies are shown in Appendix~A/Figures~\ref{f:cmd1} to~\ref{f:cmd11}. All the CMDs
are obtained from the version v2 of the photometric catalogs. The NUV CMDs show only stars with {\tt GPQF}$=55***$, in the 
optical CMD only stars with {\tt GPQF}$=***55$ are shown. A careful inspection of the optical CMDs for three of the larger 
spiral galaxies shows that metallicity is increasing moving from the outskirts towards the center. The most prominent example 
is NGC~5457 (a.k.a M101), in which we find that the color of the stellar populations in the most external field (NGC~5457nw3) are 
better reproduced by isochrones for Z=0.0008, isochrones with metallicity Z=0.008 are more indicated for the central pointing (NGC~5457c), 
while for the three intermediate pointing (NGC~5457se, NGC~5457nw1 and NGC~5457nw2) we used isochrones with metallicity Z=0.004. Similarly 
for NGC~628, we used Z=0.02 for the central pointing (NGC~628c), while 
Z=0.008 seams to be more appropriate for eastern pointing (NGC~628e). Similar metallicity gradients have been previously observed 
in several other spiral galaxies \citep{munoz07, macarthur09, gogarten10}, and have been interpreted as an indication that large
disks have been forming ``inside-out''. In this scenario, star formation started in the center, therefore we should find a higher
fraction of older stars in the core of galaxies compared to the outskirt. This would favor a higher metal enrichment of the interstellar medium in the center of the 
galaxy, and therefore cause the more recently formed stars toward the center to have higher metal abundances than stars formed in the 
outskirts of the disk.

The galaxies in our sample are affected by different amounts of dust extinction. For each galaxy we used the foreground 
extinction derived by \citet{schlafly11} and the \citet{fitzpatric99} reddening law R$_V$=3.1. A comparison of the  (NUV-U) 
and (V-I) colors with PARSEC isochrones shows that in most of the cases the upper MS is affected by a higher amount of 
extinction than just the Galactic foreground extinction. This is consistent with the fact that stars tend to form in 
dust-rich environments. 

The properties of the galaxies studied in the LEGUS project are summarized in Table~\ref{t:gal_par1}, including the foreground 
E(B-V), listed in column~7, and the E(B-V) values derived from the comparison of the upper MS colors with the models are 
reported in column~8. In addition the Table reports the name of the galaxy in column~1, the RC3 morphological {\it T}-type is 
listed in column~2; stellar and H{\sc i} masses (expressed in $M_\odot$) in colums~3  and 4 respectively. Column~5 reports the 
SFR corrected for dust attenuation derived from the Galex FUV flux assuming a Kroupa initial mass function, and column~6 reports 
the FUV half light radius R$_{1/2}$. The metallicity Z as inferred from the comparison of the optical CMDs with PARSEC 
isochrones is reported in column 9. Distance moduli and distances with their respective uncertainties are reported on 
columns~10 to 13 (see discussion in Section~\ref{trgb}). Column~14 reports the number of stars more massive that 14 $M_\odot$ 
found in the field of each galaxy (see discussion in Section~\ref{gamma}).

The optical CMDs have been corrected for distance, as derived from the luminosity of the TRGB (see Section~\ref{trgb}), and foreground 
extinction. Since the majority of the stars in the NUV CMDs are in the MS evolutionary phase, the NUV CMDs have been corrected 
for the extra amount of extinction derived from the MS colors. The blue-dotted line in the NUV CMDs marks the luminosity of 
a $M=14 M_\odot$ star in the NUV filter for the assumed metallicity, while the dashed line in the optical CMDs highlights the 
position of the TRGB, as derived in the following section. The galaxies are shown in order of increasing metallicity and SFR. 

\section{The tip of the RGB and galaxy distance moduli}
\label{trgb}

The explosive onset of helium burning in the degenerate core of low-mass stars drastically ends the RGB evolutionary phase. 
At low metallicity this occurs at about the same brightness for all low-mass stars, and, in the I vs V-I CMD, causes a sharp 
cutoff in the star counts, the so called TRGB. The luminosity of the TRGB is a standard candle and it has been used to 
determine the distances of galaxies for more than 35 years \citep{iben83, da costa90, lee93, sakai96}. Distances derived from 
the luminosity of the TRGB agree with those inferred from the Cepheid period-luminosity relation at the level of 0.01 mag 
\citep{rizzi07}. 

Several studies have demonstrated that in order to obtain a robust determination of the TRGB, it should be at least one magnitude 
above the detection threshold \citep[e.g.][]{madore95, makarov06, madore09}. So far the TRGB has been used to determine the 
distance of galaxies closer than $~$18--20 Mpc \citep{aloisi07, tully16} 

For low metallicities ($[Fe/H]<-0.7$) the luminosity of the TRGB in the I band is independent of the age of the stellar 
population, and the effect of metallicity is less than 0.2 mag \citep{lee93, salaris97, bellazzini01}. At higher metallicity 
the luminosity of the TRGB depends on the age of the stellar population. \citet{rizzi07} have shown that this degeneracy can 
be broken by using the color of the RGB. In our analysis we decided to adopt a more conservative approach and leave the 
metallicity as an unknown factor. By averaging over the metallicity range and using PARSEC isochrones, we found that in the 
WFC3 Vegamag photometric system the absolute magnitude of the TRGB is $M_I=-4.0$, with a 0.2 uncertainty due to the lack of 
information on metallicity to be added in quadrature to the photometric and reddening errors.   

When possible, we measured the apparent magnitude of the TRGB in the outskirts of the galaxies, where the ``contamination''  
from the young and intermediate-age stellar populations is negligible. In those cases where it was not possible to identify 
a region in the galaxy dominated by the old (age $>1$ Gyr) stellar population, we removed all the sources that were detected 
in the three bluer filters of our survey. 

We applied an edge-detection algorithm based on Sobel's filter to the Gaussian-smoothed I luminosity function, following the 
prescription of \citet{sakai96}. Because extinction tends to be lower in the galaxy outskirts, for each galaxy the luminosity 
of the TRGB was corrected for the foreground extinction only.

The accuracy of the TRGB detection using the Sobel filter depends on the step used for the binning. We used a 0.05 magnitude 
step and a 0.01 sigma for the smoothing. The small binning step increases the numbers of peaks identified by Sobel's 
filter and it is necessary to introduce a likelihood method to identify the peak corresponding to the TRGB. We ran simulations 
using PARSEC isochrones for the metallicity estimated in the previous section, the photometric errors at the various magnitudes 
and colors, the foreground extinction, and the magnitude of the candidate TRGBs. 

For the larger galaxies, where multiple pointings were available, we independently derived the distance modulus for each field, 
and then used the average value for the distance of the galaxy. In most cases our measurements are in excellent agreement 
($\Delta(m-M)_0<0.1$) with the values found in the literature \citep[e.g.,][]{tosi01, jacobs09, tully13, skillman13, sacchi16}. 
However it should be noted that in the central pointing of NGC~1512 crowding was so severe that it was not possible to 
determine the luminosity of the TRGB from the data. In this case we assumed the average values estimated from the SW pointing 
and NGC~1510. 

Crowding was an issue also in the detection of the TRGB in the center of NGC~628 and NGC~5194. For these two regions we 
assumed the distance derived from the other fields. A combination of distance, crowding, and metallicity made the estimate of 
the TRGB luminosity of the galaxies NGC~1433, NGC~5195, NGC~1512, NGC~3351, and NGC~3627 quite uncertain, and these values 
should be used with caution. The TRGB of IC~559, NGC~2500, NGC~1510, NGC4594, and NGC3368 is less than a magnitude above the 
detection threshold, thus the distance for these galaxies could have been underestimated. Finally, NGC~1566 is too far away to 
detect the tip in our photometry. For this galaxy we assumed a distance of $18\pm 2$ Mpc (Tully, private communication). 

\section{On the Clustering Properties of Star Formation}
\label{gamma}

\begin{figure}
\epsscale{1.0}
\plotone{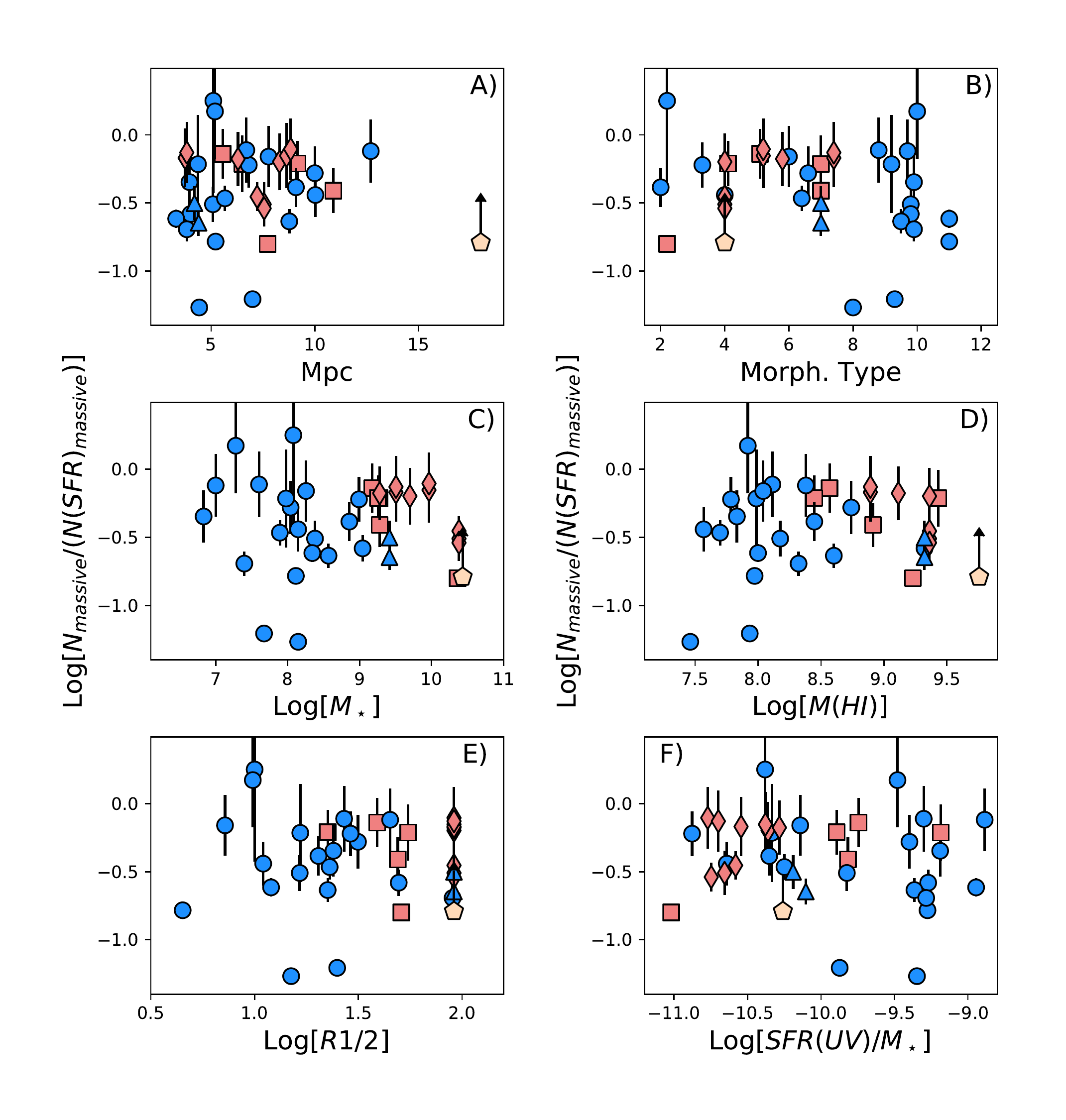}
\caption{\label{f:sample} {\it Panel A)}: Ratio between the number of stars above 14~M$_{\odot}$ as counted from the LEGUS 
data and the expected number of stars more massive than 14~M$_{\odot}$ as derived from the measured SFRs as a function of 
distance ({\it Panel A)}), morphological type  ({\it Panel B)}), total stellar mass  ({\it Panel C)}), mass of gas ({\it Panel 
D)}), UV half light radius  ({\it Panel E)}) and sSFR ({\it Panel F)}). Spiral galaxies are marked in orange and dwarf galaxies 
are marked in blue. Spiral galaxies that fit within the WFC3 FoV are marked with square symbols, while spirals significantly 
larger than the WFC3 FoV are marked with diamonds. Similarly, dwarf galaxies that fit in the WFC3 FoV are indicated by circles, while 
dwarfs whose half light radius is significantly larger that the WFC3's FoV are indicated by triangles. NGC~1566, for which 
we have only a lower limit, is indicated by a yellow pentagon.}
\end{figure}

The NUV vs NUV-U CMDs can be used to identify the young massive stars (i.e. $M\ge 14\ M_\odot$) present in the field of a galaxy. 
Because the majority of the massive stars are still likely in star clusters and stellar associations, and therefore cannot be resolved 
into single stellar objects, these sources will be not accounted for in the NUV vs NUV-U CMDs. The galaxy's FUV luminosity, however, 
includes the contribution coming from all the stars, whether they are in clusters, associations, and in the field ($L_{FUV}=L_{field\ stars}+L_{young\ bound\ 
clusters}+L_{young\ unbound\ associations}$). Therefore we can infer the fraction of objects that are still in clustered environments 
few Myr after their formation (i.e., within $\simeq 14$ Myr from birth) by comparing the number of apparently single stars found in the 
field with the number of massive stars expected from the galaxy's FUV luminosity.

We used PARSEC isochrones of appropriate metallicities to determine the absolute NUV magnitude of the turn-off (TO) of a 14 
Myr old stellar population. This corresponds to a $14\, {\rm M_\odot}$ star. For each galaxy we then counted how many stars 
are brighter than the 14 Myr TO. 

To derive the number of expected massive stars, we started by calculating the SFRs for the LEGUS galaxies (or, for large 
galaxies, for regions within the galaxy) from the combination of FUV and 24~$\mu$m emission from GALEX and Spitzer/MIPS, 
respectively, using the equation of \citet{hao11}, and assuming the distances derived from the luminosity of the TRGB 
(column 12 of Table~\ref{t:gal_par1}). We then normalize the SFRs to the \citet{kroupa01} IMF in the stellar mass 
range 0.1-100~M$_{\odot}$. The combination of observed UV and 24~$\mu$m accounts for the presence of dust in the regions, 
and provides an extinction--corrected SFR. The use of the UV emission provides mean SFRs over the most recent $\sim$100~Myr 
timescale \citep{kennicutt12, calzetti13}.

In order to derive a SFR surface density $\Sigma_{SFR}$, SFRs were normalized to the galaxy areas, calculated from the GALEX 
FUV half--light radius R$_{1/2}$ (Table~\ref{t:gal_par1}, column 5), assuming the distance derived from the TRGB. 
We measure R$_{1/2}$ directly from the GALEX archival images, using a circular aperture. 

With the exception of NGC~628, NGC~1313, NGC~3344, NGC~1566, NGC~3344, NGC~5194, NGC~6503, and NGC~7793, for which we derived 
the SFR from the UV$+$24$\mu$m light within the WFC3 field of view, galaxies with a UV R$_{1/2}> 55^{\prime\prime}$ were 
excluded from the study presented in this section, as a significant fraction of their SF (and massive stars) are outside the 
footprint of the WFC3/UVIS camera. For four of the galaxies with multiple pointings, we derived the SFR specific to the pointing. 
These galaxies are: NGC~628, NGC~1313, NGC~5194, and NGC~7793. In total we derived SFRs for 32 galaxies.

We used the SFRs obtained above to derive the expected numbers of stars more massive than 14~M$_{\odot}$. Since the SFRs 
include a timescale, we used a luminosity--weighted timescale, which, for 14~M$_{\odot}$, a Kroupa IMF, and stellar 
mass--luminosity relations as given in \citet{gallagher00}, is about 4$\times$10$^6$~yr. In all our calculations, we assumed 
that stars below 2~M$_{\odot}$ provide negligible contribution to the UV luminosity, and that the SFR remained constant over 
the last 100~Myr. The last assumption is likely our largest source of uncertainty in the numbers we obtained, especially for 
the dwarf galaxies, because they tend to form stars in bursting episodes.

For the 32 galaxies mentioned above we derived the ratio: N$_{massive}$/N(SFR)$_{massive}$, which is the ratio between the 
number of stars above the 14~Myr old TO, counted in the LEGUS data, and the number of stars more massive that 
14~M$_\odot$, as expected from the SFRs derived from GALEX. We should note that, within the uncertainties, this ratio 
is expected to be always $\le$1, since the resolved stars do not include either star clusters or compact associations. However 
for UGCA~281 and IC~4247 we find that N$_{massive}$/N(SFR)$_{massive}\ge1$. This is probably an artifact due to the assumption 
of a constant SFR over the past 100 Myr. Both IC~4247 and UGCA~892 in fact are small dwarf irregular galaxies, that experienced 
very modest SF in the past few Myr.

Figure~\ref{f:sample} shows how the ratio N$_{massive}$/N(SFR)$_{massive}$ relates to a number of global properties of the 
galaxies, such as distance ({\it Panel A)}), RC3 morphological {\it T}-type as discussed in \citet[][ {\it Panel B)}]{kennicutt08}, 
stellar mass ({\it Panel C)}), total mass in neutral gas, FUV effective radius, and specific SFR (sSFR=SFR/mass, {\it Panel D)}). 
Spiral galaxies are marked in orange, using either square symbols (when the entire NUV half-light falls within the WFC3 FoV), or
diamonds (when their SFRs have been normalized to the WFC3 FoV). Dwarf galaxies, marked in blue, are indicated with circles (when they fit
within WFC3 FoV), or triangles. This figure is used to establish whether our measurements may suffer from biases or selection 
effects. 

As mentioned above, we assumed constant SFR over the past 100 Myr. While this assumption is probably correct for spiral and Magellanic 
galaxies, for the dwarf systems it is more likely that SF occurred in short burst. Depending on when and at which rate SF developed in the dwarf 
galaxies, this can cause either an over or an under-estimation of the total number of massive stars. {\it Panel A)} of Figure~\ref{f:sample}, 
for example, shows a  correlation between the number of massive stars found in the field of dwarf galaxies and their distance. 
This trend is due to the fact that the majority of the late-type (RC3 T-type$\ge 8$) dwarfs in the LEGUS sample are between 4 and 6 Mpc. In the 
case of the spiral galaxies, on the contrary, we do not notice any obvious trend with the distance. We do not find any obvious correlation 
between the ratio N$_{massive}$/N(SFR)$_{massive}$ and the other quantities mentioned above, therefore we conclude that our measurements 
are mostly free from these selections effects.

\begin{figure}
\epsscale{1.0}
\plotone{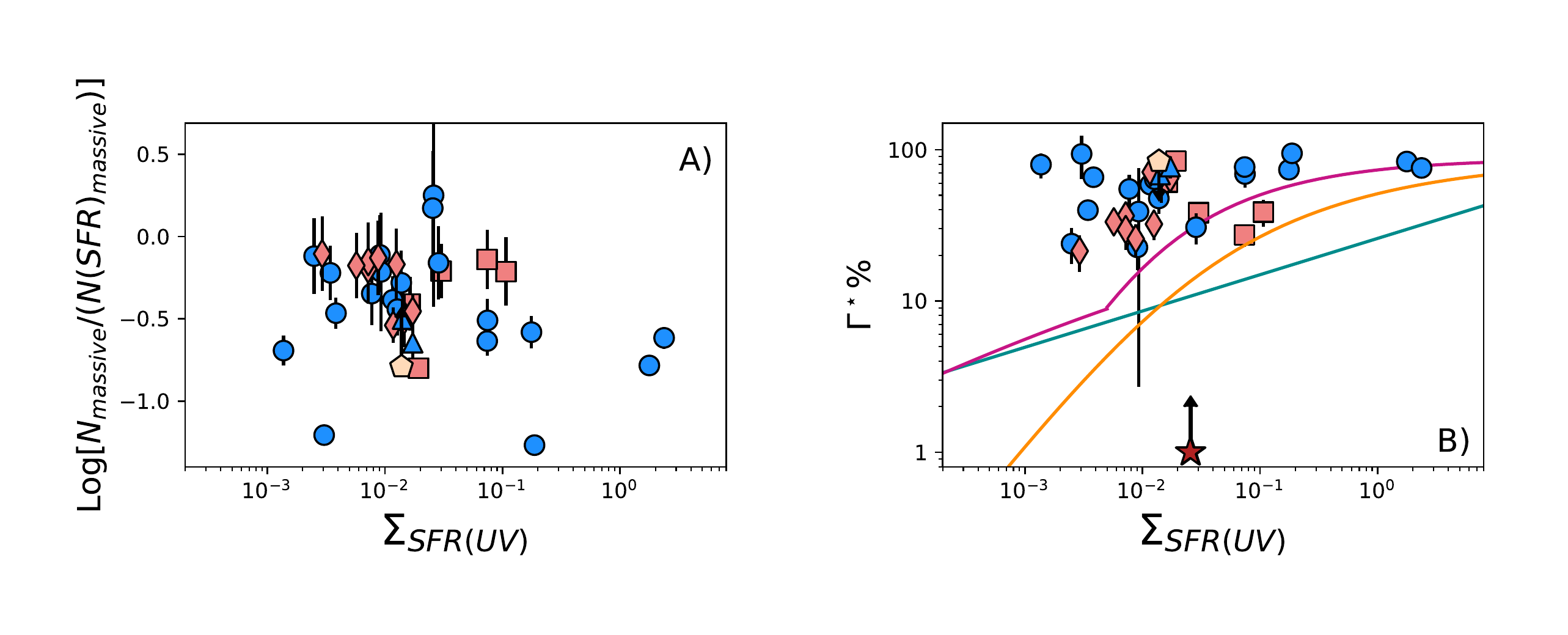}
\caption{\label{f:NC} {\it Panel A)}: Ratio between the number of stars above 14~M$_{\odot}$ as counted from the LEGUS data 
and the expected number of stars more massive than 14~M$_{\odot}$ as derived from the measured SFRs as a function of 
$\Sigma_{SFR}$. {\it Panel B)}: Fraction of stars that likely formed in clustered environments as a function of $\Sigma_{SFR}$. 
As in Figure~\ref{f:sample} spiral galaxies are in orange and dwarf galaxies are blue. Spiral galaxies that fit within the WFC3 
FoV are marked with square symbols, while for spirals significantly larger than the WFC3 FoV we used diamond symbols. 
NGC~1566, for which we have only a lower limit is marked with a yellow pentagon. Circles represent dwarf galaxies that fit within the 
WFC3 FoV, while larger dwarfs are indicated with triangles. The red stars in Panel B mark the two targets (IC~4247 and UGCA~281) whose 
number of stars counted in the field exceeds the number of stars expected from the SFR. Our data are compared to analytical predictions 
for the values of $\Gamma$, derived assuming a constant \citep[yellow line][]{kruijseen12}, and variable gas surface density
\citep[magenta line]{johnson16}, and to a fitted relation obtained from observational $Gamma$ values 
\citep[cyan line][]{goddard10}.
}
\end{figure}

{\it Panel A)} of Figure~\ref{f:NC} shows the ratio N$_{massive}$/N(SFR)$_{massive}$ as a function of $\Sigma_{SFR}$. Again, 
we do not notice any obvious trend in this plot. 



We introduce now a new value 
\begin{equation}
\Gamma^\star = 1- {N_{massive} \over N(SFR)_{massive}}
\end{equation}
as the fraction of stars that are {\em not} in the field, and, therefore, are likely to be in clustered star forming regions (including both 
compact bound clusters and unbound associations). Hence, we defined the above quantity as $\Gamma^\star$, rather than 
$\Gamma$, since the latter quantity usually refers to the fraction of stellar mass that is formed in bound clusters only. 

We should note that since we measured both N$_{massive}$ and N(SFR)$_{massive}$ for the same massive stars, their number ratio is 
equivalent to the mass ratio. Our parameter, being derived from stars more massive than 14~M$_{\odot}$, probes timescales shorter 
than 14~Myr, and is unlikely to measure only the fraction of stellar mass in bound systems. Rather, it is measuring the fraction of 
stellar mass that is included in all young clusters and associations. Still, because our derivation covers the same restricted age-range 
for all the galaxies and regions considered, over the three orders of magnitude-range of $\Sigma_{SFR}$, it enables several considerations.
 
Both observations and models indicate that the fraction of star formation occurring in bound stellar clusters changes as a function 
of $\Sigma_{SFR}$. To probe whether $\Gamma^\star$ scales similarly as $\Gamma$, we plot in {\it Panel B)} of Figure~\ref{f:NC},  
$\Gamma^\star$ as a function of $\Sigma_{SFR}$. The plot includes the prediction by \citet[][yellow line]{kruijseen12}, a modified 
version of the Kruijssen model that takes into account the change in the slope of the relation between $\Sigma_{SFR}$ and $\Sigma_{gas}$ 
\citep[][magenta line]{johnson16}, an observational fit to the observed distributions in the $\Gamma$ versus $\Sigma_{SFR}$ plane proposed 
by \citet[][cyan line]{goddard10}. Although there is scatter in the value of $Gamma$ at different $\Sigma_{SFR}$, we assume that these 
models and fit identify the loci occupied by the formation efficiencies of bound cluster populations as a function of the $\Sigma_{SFR}$ 
of the host galaxies.

From the right panel of Figure~\ref{f:NC}, we can readily see that:
\begin{itemize}
\item The observed values of $\Gamma^\star$ cover almost one order of magnitude at the low end of our range of $\Sigma_{SFR}$. 
\item Most of the data are above the predictions of all the bound-cluster curves, with $\sim 14\%$ points having $\Gamma^\star >80\%$. 
This suggests that in these regions most of the stars form in clustered environments (either bound or unbound) and remain close to their 
siblings for at least the first 10-15 Myr. 
\item For $\Sigma_{SFR}<0.1$, $50\%$ of the points have $\Gamma^\star$ below 50\%.
\item For intermediate values of $\Sigma_{SFR}$, $\Gamma^\star$ is close and even above the values predicted by \citet{kruijseen12}, and 
\citet{johnson16}. 
\item Although there is a large scatter, the $\Gamma^\star$ values at $\Sigma_{SFR} <0.01$  are systematically higher than the predictions 
from the models. 
\item Because $\Gamma^\star$ includes the fraction of both bound and unbound systems, this may indicate that the timescale required for 
association dissolution is longer than the usually-adopted 10 Myr \citep{fall05}. 
\end{itemize}

The uncertainties in the assumption of a constant SFR over the past 100~Myr are likely to be about a factor 3, which is smaller 
than the scatter in the $\Gamma^\star$ values for $\Sigma_{SFR}\la 0.1$. Furthermore, some of the 
data with the lowest $\Gamma^\star$ values are regions within large galaxies, where we expect the star formation history to 
have remained relatively constant. Thus, at the low end of the $\Sigma_{SFR}$ values, the fraction of stars that are in 
clustered structures spans almost the full range of $\Gamma^\star$ values, from less than 20\% to almost 100\%. This scatter 
is unlikely to be due to spurious factors like distance--dependency in our ability to resolve stars, since there is no dependency of 
N$_{massive}$/N(SFR)$_{massive}$ on distance (Figure~\ref{t:gal_par1} - {\it Panel A)}). 

Conversely, at the high end of the $\Sigma_{SFR}$ range, we observe a decrease in the scatter of $\Gamma^\star$. Although low 
number statistics may have an effect on this trend, it is still informative that all data are around or slightly above 
predictions from models. Furthermore, the scatter in $\Gamma^\star$ for $\Sigma_{SFR}\ga 0.1$ is about a factor 30 lower than 
for $\Sigma_{SFR}\la 0.1$. This appears to indicate that high $\Sigma_{SFR}$ galaxies, when they form clusters, retain a higher 
fraction of bound systems, thus the fraction contributing to the field (i.e.,dissolving) is proportionally a smaller fraction of 
the total SFR.

Roughly, the models we report in the right panel of Figure~\ref{f:NC} mark a lower boundary to the values of $\Gamma^\star$ we 
measure, suggesting that stars form preferentially in clustered environments and remain confined within these compact (both bound and 
unbound) structures for the first $\sim$10-15~Myr. This result implies that $\Gamma$ (cluster formation efficiency of bound cluster) 
derived using ages below $\sim$10 Myr and cluster catalogues based on the compactness of systems, as for the LEGUS survey, will always 
suffer of contamination of systems that are in reality unbound (e.g. Adamo et al. 2017). Indeed, \citep{messa18} reports that in M51 
$\Gamma$ estimated in the age range 1-10 Myr is a factor of 3 higher than if estimated in the age range 10-100 Myr, i.e. well within 
the scatter observed with our analysis of $\Gamma^\star$.
 

\begin{landscape}
\begin{deluxetable}{lrrrrrrrrrrrrr}
\tablecolumns{14}
\tablewidth{0pt}
\tabletypesize{\tiny}
\tablecaption{\label{t:gal_par1} Properties of the LEGUS galaxies. Galaxies for which the derived distance is uncertain have been marked with an asterisk or a \^\ symbol. 
The galaxies used to evaluate the relation between $\Gamma^\star$ and the $\Sigma_{SFR}$ are highlighted in boldface.} 
\tablehead{
\colhead{Name} & \colhead{{\it T}-type} & \colhead{M$_\star$} & \colhead{M(H{\sc i})} & \colhead{SFR(UV)} & \colhead{FUV R$_{1/2}$} & \colhead{E(B-V)$_0$} & \colhead{E(B-V)} & 
\colhead{Z} & \colhead{(m-M)$_0$} & \colhead{Err (m-M)$_0$} & \colhead{Dist} & \colhead{Err Dist} & \colhead{N$_{14M_\odot}$} 
\\
\colhead{}  & \colhead{} & \colhead{($M_\odot$)} & \colhead{($M_\odot$)} & \colhead{($M_\odot$ year$^{-1}$)} & \colhead{(arcsec)} & \colhead{} & \colhead{} & \colhead{} & \colhead{} & \colhead{} & \colhead{Mpc} & \colhead{Mpc} & \colhead{}}
\startdata

{\bf ESO4486}	      &    2.0    &   7.2e8   &  2.8e8   &  0.0306   &  20.3  & 0.03  & 0.04   & 0.004	  &   29.79  &  0.33   &   9.09	 &  0.7	    &  198 \\
{\bf IC~4247}	      &    2.2    &   1.2e8   &  1.0e1   &  0.005    &  10.0  & 0.06  & 0.06   & 0.0008	  &   28.54  &  0.32   &   5.11	 &  0.4	    &  146 \\
{\bf IC~559}\^      &    4.0    &   1.4e8   &  3.7e7   &  0.0061   &  11.0  & 0.02  & 0.04   & 0.004	  &   30.00  &  0.37   &  10.0   &  0.9     &   68 \\
NGC~1291n	  &    0.1    &   1.5e11  &  2.3e9   &  0.2344	 &   0.0  & 0.01  & 0.16   & 0.02	  &   28.98  &  0.34   &   6.3	 &  0.5	    &  120 \\
{\bf NGC~1313e}	  &    7.0    &   2.6e9   &  2.1e9   &  0.1607   &  91.4  & 0.08  & 0.12   & 0.008	  &   28.11  &  0.32   &   4.2	 &  0.34    &  794 \\
{\bf NGC~1313w}	  &    7.0    &   2.6e9   &  2.1e9   &  0.2055   &  91.4  & 0.08  & 0.12   & 0.008	  &   28.22  &  0.32   &   4.4	 &  0.35    &  763 \\
NGC~1433*	  &    1.5    &   1.7e10  &  5.0e8   &  0.1856	 &  74.0  & 0.01  & 0.03   & 0.02	  &   29.78  &  0.49   &   9.1	 &  1.0     &  187 \\
NGC~1512c*    &    1.1    &   1.7e10  &  8.5e9   &  0.6327	 &   0.0  & 0.01  & 0.02   & 0.02	  &   30.33  &  0.40   &  11.7   &  1.1     &  468 \\
NGC~1512sw1*  &    1.1    &   1.7e10  &  8.5e9   &  0.6463	 &   0.0  & 0.01  & 0.03   & 0.02	  &   30.38  &  0.45   &  11.9	 &  1.3     &  130 \\
NGC~1512sw2\^ &    1.6    &   4.8e8   &  6.5e7   &  0.0731   &   4.8  & 0.01  & 0.04   & 0.008	  &   30.28  &  0.33   &  11.4	 &  0.9     &  231 \\
{\bf NGC~1566}**	  &    4.0    &   2.7e10  &  5.7e9   &  2.026    &  91.4  & 0.01  & 0.25   & 0.02	  &   31.27  &  0.49   &  18.0	 &  2.0	    & 7323 \\
{\bf NGC~1705}	  &   11.0    &   1.3e8   &  9.4e7   &  0.0706   &   4.5  & 0.01  & 0.03   & 0.008	  &   28.59  &  0.32   &   5.22	 &  0.38    &  195 \\
{\bf NGC~2500}\^	  &    7.0    &   1.9e9   &  8.2e8   &  0.3161	 &  49.0  & 0.03  & 0.3    & 0.004	  &   30.18  &  0.46   &  10.9	 &  1.2	    & 2200 \\
{\bf NGC~3274}	  &    6.6    &   1.1e8   &  5.5e8   &  0.0667   &  31.5  & 0.02  & 0.07   & 0.004	  &   30.01  &  0.39   &  10.0	 &  0.9	    &  867 \\
{\bf NGC~3344}	  &    4.0    &   5.0e9   &  2.3e9   &  0.2585   &  91.4  & 0.03  & 0.26   & 0.02	  &   29.59  &  0.37   &   8.3	 &  0.7	    & 3188 \\
NGC~3351*	  &    3.1    &   2.1e10  &  1.3e9   &  0.9114	 &   0.0  & 0.03  & 0.2    & 0.02	  &   29.84  &  0.40   &   9.3	 &  0.9	    & 1475 \\
NGC~3368\^	  &    1.9    &   4.8e10  &  2.7e9   &  0.644	 & 147.0  & 0.02  & 0.2    & 0.02	  &   29.96  &  0.56   &   9.8	 &  1.3	    &  803 \\
NGC~3627*	  &    3.1    &   3.1e10  &  1.5e9   &  3.242	 &   0.0  & 0.03  & 0.27   & 0.02	  &   30.14  &  0.54   &  10.7	 &  1.4	    & 3443 \\
{\bf NGC~3738}	  &    9.8    &   2.4e8   &  1.5e8   &  0.0374   &  16.5  & 0.01  & 0.08   & 0.004	  &   28.53  &  0.32   &   5.09	 &  0.40    &  197 \\
NGC~4242	  &    7.9    &   1.1e9   &  3.5e8   &  0.0642   &  75.0  & 0.04  & 0.05   & 0.02	  &   28.61  &  0.24   &   5.3	 &  0.3	    &   41 \\
{\bf NGC~4248}	  &    3.3    &   9.8e8   &  6.1e7   &  0.0114   &  29.0  & 0.02  & 0.25   & 0.02	  &   29.17  &  0.32   &   6.82	 &  0.51    &   98 \\
NGC~4258n	  &    4.0    &   2.9e10  &  7.3e9   &  1.357	 &   0.0  & 0.02  & 0.2    & 0.008	  &   29.17  &  0.34   &   6.83	 &  0.54    & 2192 \\
NGC~4258s	  &    4.0    &   2.9e10  &  7.3e9   &  1.357	 &   0.0  & 0.02  & 0.2    & 0.008	  &   29.17  &  0.34   &   6.83	 &  0.54    & 2896 \\
NGC~4395n	  &    8.9    &   6.0e8   &  1.8e9   &  0.2256	 &   0.0  & 0.02  & 0.03   & 0.004	  &   28.32  &  0.32   &   4.62	 &  0.2	    &  149 \\ 
NGC~4395s	  &    8.9    &   6.0e8   &  1.8e9   &  0.2154	 &   0.0  & 0.02  & 0.04   & 0.004	  &   28.22  &  0.32   &   4.41	 &  0.33    &  402 \\
{\bf NGC~4449}	  &    9.8    &   1.1e9   &  2.1e9   &  0.5502	 &  49.5  & 0.02  & 0.14   & 0.004	  &   28.02  &  0.32   &   4.01	 &  0.30    & 2206 \\
{\bf NGC~4485}	  &    9.5    &   3.7e8   &  4.0e8   &  0.1846	 &  22.5  & 0.02  & 0.11   & 0.004	  &   29.71  &  0.35   &   8.8	 &  0.7     &  809 \\
{\bf NGC~4490}	  &    7.0    &   1.9e9   &  2.7e9   &  1.119	 &  55.0  & 0.02  & 0.25   & 0.004    &   29.06  &  0.35   &   6.5	 &  0.5     & 10148 \\
NGC~45  	  &    7.8    &   3.3e9   &  2.5e9   &  0.2263	 & 154.5  & 0.02  & 0.04   & 0.004	  &   29.16  &  0.36   &   6.8	 &  0.5     &  201 \\
NGC~4594\^	  &    1.1    &   1.5e11  &  2.8e8   &  0.5502	 &   0.0  & 0.01  & 0.22   & 0.02	  &   29.99  &  0.47   &   9.9	 &  1.1     &   36 \\
{\bf NGC~4605}	  &    5.1    &   1.5e9   &  3.7e8   &  0.2634	 &  39.0  & 0.01  & 0.28   & 0.004	  &   28.72  &  0.34   &   5.56	 &  0.44    & 3059 \\
NGC~4656	  &    9.9    &   4.0e8   &  2.2e9   &  0.443	 & 173.0  & 0.01  & 0.07   & 0.004	  &   29.48  &  0.37   &   7.9	 &  0.7     & 1251 \\
{\bf NGC~5194c}*	  &    4.0    &   2.4e10  &  2.3e9   &  0.5204	 &  91.4  & 0.03  & 0.26   & 0.02	  &   29.39  &  0.47   &   7.6	 &  0.8     & 2590 \\
{\bf NGC~5194ne}	  &    4.0    &   2.4e10  &  2.3e9   &  0.5907	 &  91.4  & 0.03  & 0.29   & 0.02	  &   29.29  &  0.33   &   7.2	 &  0.6     & 3192 \\
{\bf NGC~5194sw}	  &    4.0    &   2.4e10  &  2.3e9   &  0.4222	 &  91.4  & 0.03  & 0.23   & 0.02	  &   29.39  &  0.35   &   7.6	 &  0.6     & 1961 \\
{\bf NGC~5195*}	  &    2.2    &   2.3e10  &  1.7e9   &  0.2209	 &  51.0  & 0.03  & 0.39   & 0.02	  &   29.44  &  0.35   &   7.7	 &  0.6     &  576 \\
{\bf NGC~5238}	  &    8.0    &   1.4e8   &  2.9e7   &  0.062    &  15.0  & 0.01  & 0.03   & 0.0008	  &   28.23  &  0.32   &   4.43	 &  0.34    &   54 \\
{\bf NGC~5253}	  &   11.0    &   2.2e8   &  1.0e8   &  0.2635	 &  12.0  & 0.05  & 0.16   & 0.0008	  &   27.60  &  0.32   &   3.32	 &  0.25    & 1103 \\
NGC~5457c     &    6.0    &   1.9e10  &  1.9e10  &  3.843	 &   0.0  & 0.01  & 0.26   & 0.008	  &   28.94  &  0.34   &   6.13	 &  0.49    & 2246 \\
NGC~5457nw1   &    6.0    &   1.9e10  &  1.9e10  &  4.513    &   0.0  & 0.01  & 0.16   & 0.008	  &   29.29  &  0.33   &   7.2	 &  0.5     &  562 \\
NGC~5457nw2	  &    6.0    &   1.9e10  &  1.9e10  &  4.213    &   0.0  & 0.01  & 0.16   & 0.004	  &   29.14  &  0.32   &   6.7	 &  0.5	    &  257 \\
NGC~5457nw3	  &    6.0    &   1.9e10  &  1.9e10  &  4.213    &   0.0  & 0.01  & 0.16   & 0.0008	  &   29.14  &  0.32   &   6.7	 &  0.5	    &  378 \\
NGC~5457se	  &    6.0    &   1.9e10  &  1.9e10  &  4.112    &   0.0  & 0.01  & 0.29   & 0.008	  &   29.09  &  0.34   &   6.6	 &  0.5     & 1636 \\
NGC~5474	  &    6.8    &   8.1e8   &  1.3e9   &  0.1638	 &  85.5  & 0.01  & 0.23   & 0.004	  &   29.08  &  0.32   &   6.6	 &  0.5     & 1460 \\
{\bf NGC~5477}	  &    8.8    &   4.0e7   &  1.3e8   &  0.0209	 &  27.0  & 0.01  & 0.04   & 0.0008	  &   29.13  &  0.32   &   6.7	 &  0.5     &  278 \\ 
{\bf NGC~5949}	  &    4.1    &   1.8e9   &  2.8e8   &  0.1475	 &  22.5  & 0.02  & 0.26   & 0.02	  &   29.81  &  0.34   &   9.2	 &  0.7     &  955 \\
{\bf NGC~628c}*	  &    5.2    &   9.2e9   &  9.2e9   &  0.3577   &  91.4  & 0.06  & 0.34   & 0.02	  &   29.68  &  0.41   &   8.6	 &  0.9     & 3833 \\
{\bf NGC~628e}	  &    5.2    &   9.2e9   &  9.2e9   &  0.1483   &  91.4  & 0.06  & 0.24   & 0.008	  &   29.73  &  0.34   &   8.8	 &  0.7     & 1816 \\
{\bf NGC~6503}	  &    5.8    &   1.9e9   &  1.3e9   &  0.1177   &  91.4  & 0.03  & 0.26   & 0.02	  &   28.99  &  0.33   &   6.3	 &  0.5     & 1528 \\
NGC~6744c     &    4.0    &   2.2e10  &  1.2e10  &  3.999	 &   0.0  & 0.04  & 0.13   & 0.02	  &   29.23  &  0.40   &   7.0	 &  0.53    &  102 \\
NGC~6744n	  &    4.0    &   2.2e10  &  1.2e10  &  5.031	 &   0.0  & 0.04  & 0.19   & 0.02	  &   29.73  &  0.39   &   8.8 	 &  0.8     & 1092 \\
{\bf NGC~7793e}	  &    7.4    &   3.2e9   &  7.8e8   &  0.0998   &  91.4  & 0.02  & 0.16   & 0.008	  &   27.87  &  0.33   &   3.75	 &  0.28    & 1208 \\
{\bf NGC~7793w}	  &    7.4    &   3.2e9   &  7.8e8   &  0.0714   &  91.4  & 0.02  & 0.14   & 0.008	  &   27.92  &  0.33   &   3.83	 &  0.29    &  967 \\
UGC~1249	  &    8.9    &   5.5e8   &  9.9e8   &  0.0866   &  97.5  & 0.07  & 0.19   & 0.004	  &   29.02  &  0.32   &   6.4	 &  0.5     &  931 \\
UGC~4305	  &    9.9    &   2.3e8   &  7.3e8   &  0.0816   & 118.0  & 0.03  & 0.06   & 0.008	  &   27.55  &  0.32   &   3.32	 &  0.25    &   58 \\
{\bf UGC~4459}	  &    9.9    &   6.8e6   &  6.8e7   &  0.0048   &  24.0  & 0.03  & 0.08   & 0.0008	  &   27.99  &  0.32   &   3.96	 &  0.30    &   38 \\
{\bf UGC~5139}	  &    9.9    &   2.5e7   &  2.1e8   &  0.0125   &  90.0  & 0.04  & 0.07   & 0.0008	  &   27.92  &  0.32   &   3.83	 &  0.29    &   40 \\
{\bf UGC~5340}	  &    9.7    &   1.0e7   &  2.4e8   &  0.028    &  45.0  & 0.02  & 0.06   & 0.0008	  &   30.52  &  0.32   &   12.7	 &  1.0     &  751 \\
{\bf UGC~685} 	  &    9.2    &   9.5e7   &  9.7e7   &  0.004    &  16.6  & 0.05  & 0.19   & 0.0008	  &   28.20  &  0.32   &   4.37	 &  0.34    &   36 \\
{\bf UGC~695} 	  &    6.0    &   1.8e8   &  1.1e8   &  0.0093   &   7.2  & 0.03  & 0.2    & 0.0008   &   29.45  &  0.32   &   7.8	 &  0.6     &   75 \\
{\bf UGC~7242}	  &    6.4    &   7.8e7   &  5.0e7   &  0.0046   &  23.0  & 0.02  & 0.03   & 0.004	  &   28.77  &  0.33   &   5.67	 &  0.43    &   27 \\
{\bf UGC~7408}	  &    9.3    &   4.7e7   &  8.6e7   &  0.0066   &  25.0  & 0.01  & 0.04   & 0.0008	  &   29.23  &  0.33   &   7.0	 &  0.5     &    7 \\
{\bf UGCA~281}	  &   10.0    &   1.9e7   &  8.3e7   &  0.0055   &   9.8  & 0.01  & 0.14   & 0.0008	  &   28.58  &  0.32   &   5.19	 &  0.39    &  119 \\
\enddata
\end{deluxetable}
\end{landscape}

\section{Summary and Conclusions}
\label{conclusions}

LEGUS (GO-13364, PI Calzetti) is a multi-wavelength survey of 50 nearby star-forming galaxies. In this paper we presented 
the data reduction and the photometric analysis of the resolved stellar populations, found in the galaxy fields. The stellar
photometry was measured by PSF fitting using the photometry package DOLPHOT by analyzing each filter independently.

As for all the HST Treasury programs, all the data collected by the LEGUS project were made immediately available to the 
astronomical community. To increase the legacy value of the project, ACS archival data in the filters B, V,  
and/or I have been aligned to the LEGUS data and can be downloaded from the LEGUS 
webpage\footnote{https://archive.stsci.edu/prepds/legus/dataproducts-public.html}. 
We are now releasing the astrophotometric catalogs for the resolved stellar populations. Cluster catalogs for most of the LEGUS
galaxies are also becoming available \citep{adamo17}. 

The NUV CMDs probe the stellar content of the LEGUS galaxies down to $M\la 5-10,\, {\rm M_\odot}$ (depending on the galaxy 
distance), with a look back times of several tens of  Myr. At the optical wavelengths the lookback time is larger than at 
least one billion years. The LEGUS galaxies cover a broad range of metallicities, and SFRs, with the larger systems showing 
evidence for metallicity and SFR gradients.

For most of the galaxies we were able to clearly identify the TRGB and derive an independent estimate of their distance. 
We used the metallicity and reddening derived by comparing the CMDs with PARSEC isochrones to estimate the number of stars 
more massive than $M=14\, {\rm M_\odot}$ in the NUV CMDs. 

By comparing the number of stars more massive that 14 $M_\odot$ found in the field of the galaxies with the expected number 
of massive stars inferred from the FUV SFR, we were able to estimate the fraction of massive stars found in clustered environments  
(bound star clusters $+$ unbound stellar associations) at early ages ($\le14$ Myr). Our survey spans 3 orders of magnitude in SFR 
density and shows that at early ages ($\leq$ 10-15 Myr) the formation efficiency of compact (bound and unbound) stellar systems ( 
$\Gamma^\star$) remains above the predicted cluster formation efficiency of bound systems and does not depend on $\Sigma_{SFR}$. We 
observe significantly more scatter in the $\Gamma^\star$ versus $\Sigma_{SFR}$ for $\Sigma_{SFR}\la 0.1$ than for higher values. We 
suggest that the reduced scatter may be driven by the increase of star formation happening in bound clusters for increasing $\Sigma_{SFR}$.
The lack of relation between $\Gamma^\star$ and $\Sigma_{SFR}$ suggests that the time scale for evaporation of unbound structures is 
comparable or longer than 10 Myr, thus $\Gamma$ estimated with cluster catalogues relying only on the compactness of the systems will be 
overpredicted if limited to short age ranges (1-10 Myr), because contaminated by compact unbound associations.

\acknowledgments
We thank the referee Nate Bastian for the constructive suggestions that considerably improved the paper.

A.A. acknowledges partial support from the Swedish Royal Academy. G.A. acknowledges support from the Science and Technology 
Facilities Council (ST/L00075X/1 and ST/M503472/1). C.D. acknowledges funding from the FP7 ERC starting grant LOCALSTAR 
(no. 280104). M.F. acknowledges support by the Science and Technology Facilities Council (grant number ST/L00075X/1). 
D.A.G. kindly acknowledges financial support by the German Research Foundation (DFG) through program GO1659/3-2. 
These observations are associated with program \# 13364. Support for program \# 13364 was provided by NASA through a grant 
from the Space Telescope Science Institute. Based on observations obtained with the NASA/ESA Hubble Space Telescope, at the 
Space Telescope Science Institute, which is operated by the Association of Universities for Research in Astronomy, Inc., under 
NASA contract NAS 5-26555.

Facility: HST (WFC3, ACS).

 \bibliographystyle{apj}

\section*{Appendix~A}

\begin{figure}
\epsscale{1.0}
\plotone{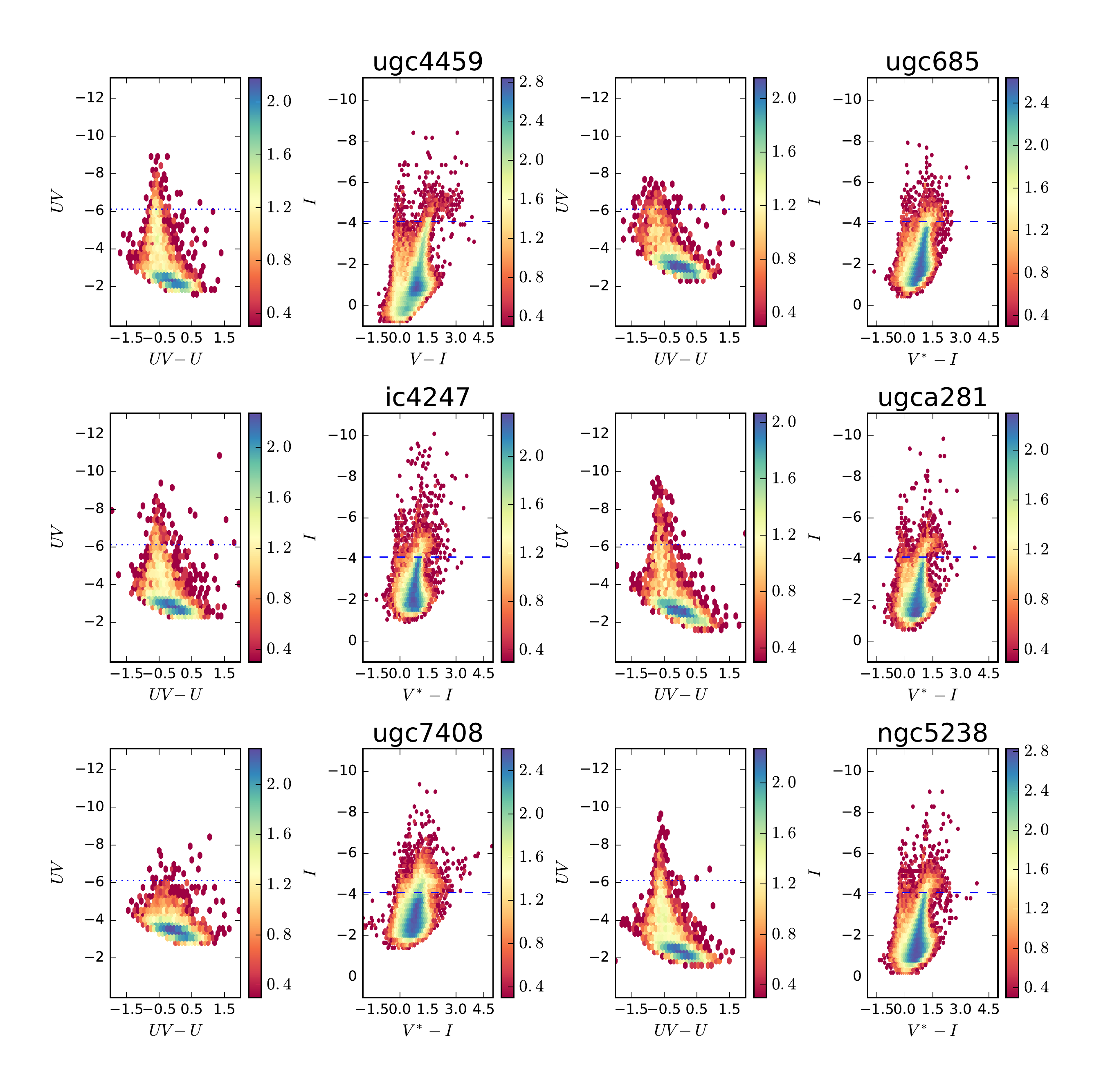}
\caption{\label{f:cmd1} NUV and optical CMDs for the galaxies UGC4459, UGC685, IC4247, UGCA281, UGC7408, and NGC5238. Magnitudes marked as V$^*$ refer
to the filter F606W, instead of F5555W.}
\end{figure}

\begin{figure}
\epsscale{1.0}
\plotone{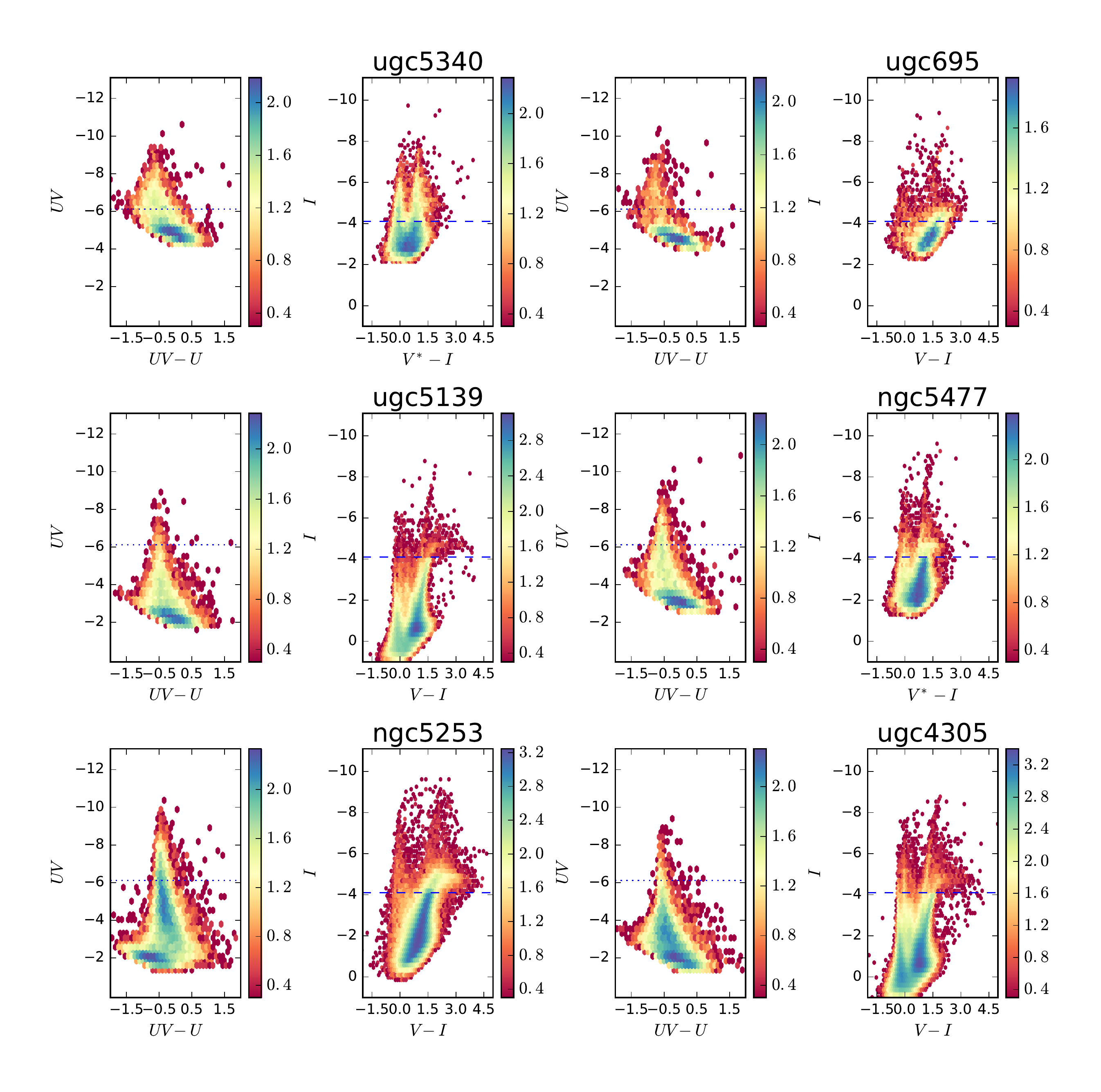}
\caption{\label{f:cmd2} Same as Figure~\ref{f:cmd1}, but for the galaxies UGC5340, UGC695, UGC5139, NGC5477, NGC5253, UGC4305. Magnitudes marked as V$^*$ refer
to the filter F606W, instead of F5555W.}
\end{figure}

\begin{figure}
\epsscale{1.0}
\plotone{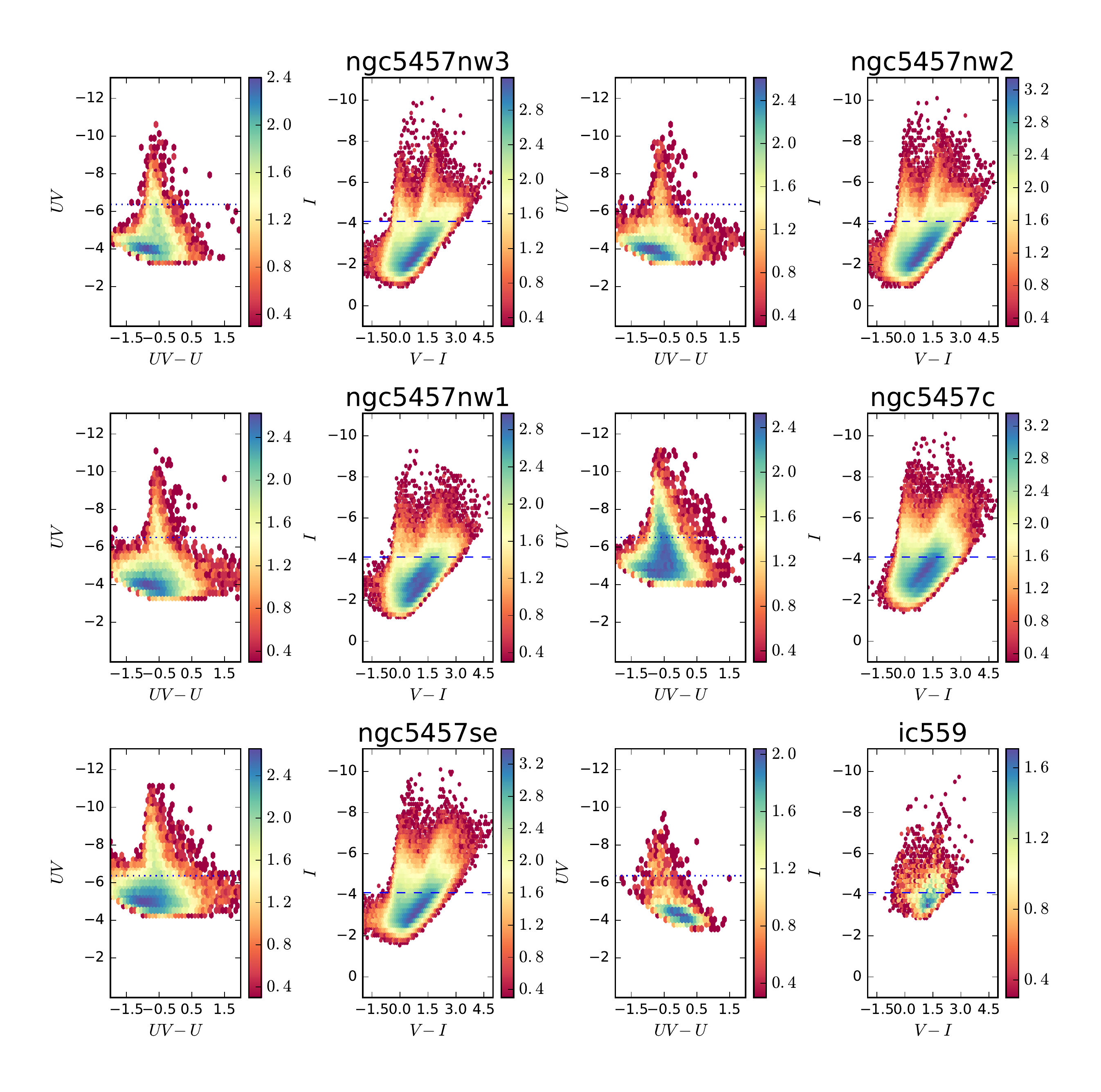}
\caption{\label{f:cmd3} Same as Figure~\ref{f:cmd1}, but for the galaxy NGC5457 (fields North West 3, 2, 1, Center and South East), and the galaxy IC559.}
\end{figure}

\begin{figure}
\epsscale{1.0}
\plotone{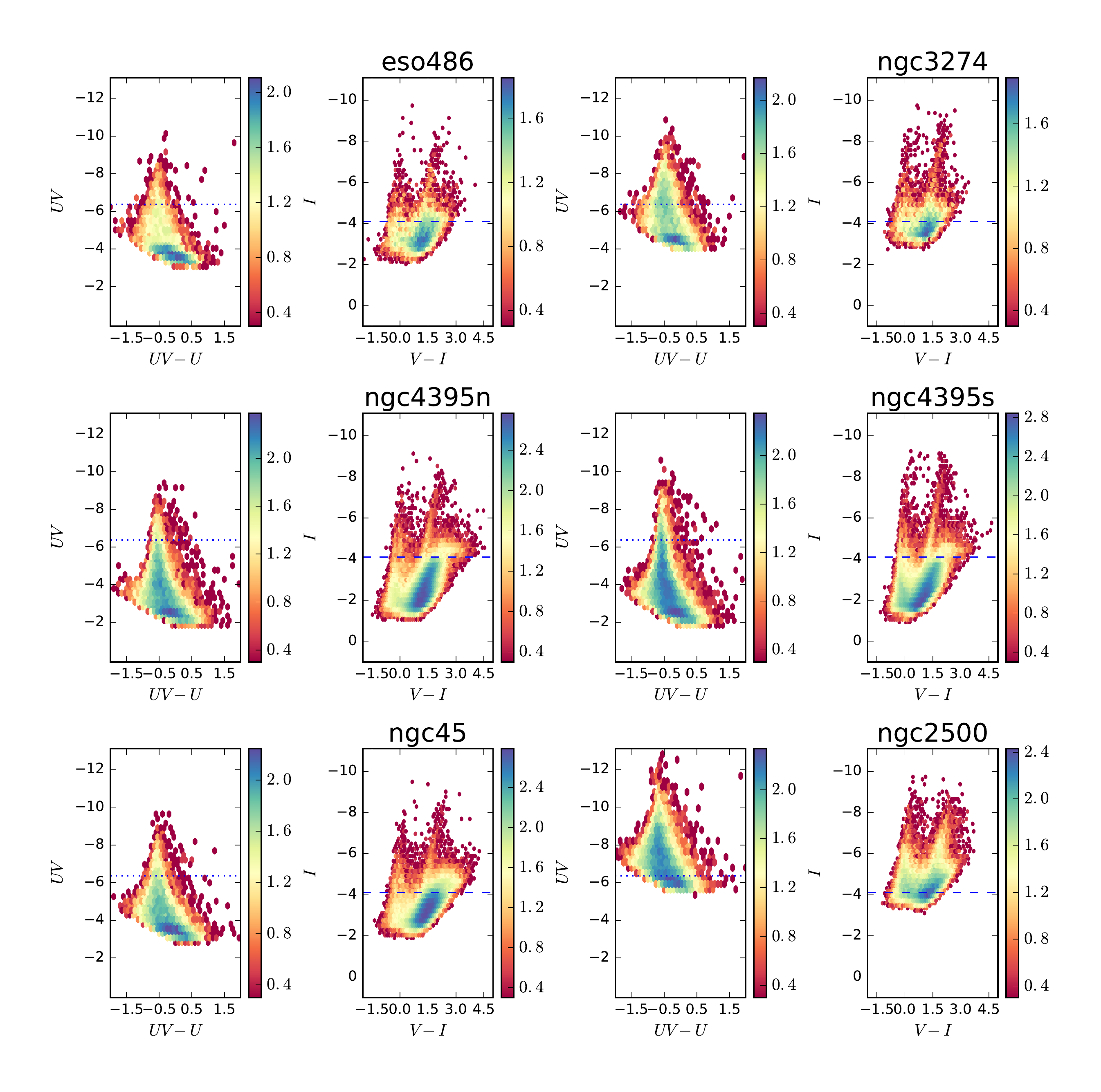}
\caption{\label{f:cmd4} Same as Figure~\ref{f:cmd1}, but for the galaxies ESO486, NGC3274, NGC4395 (North and South), NGC45, and NGC2500.}
\end{figure}

\begin{figure}
\epsscale{1.0}
\plotone{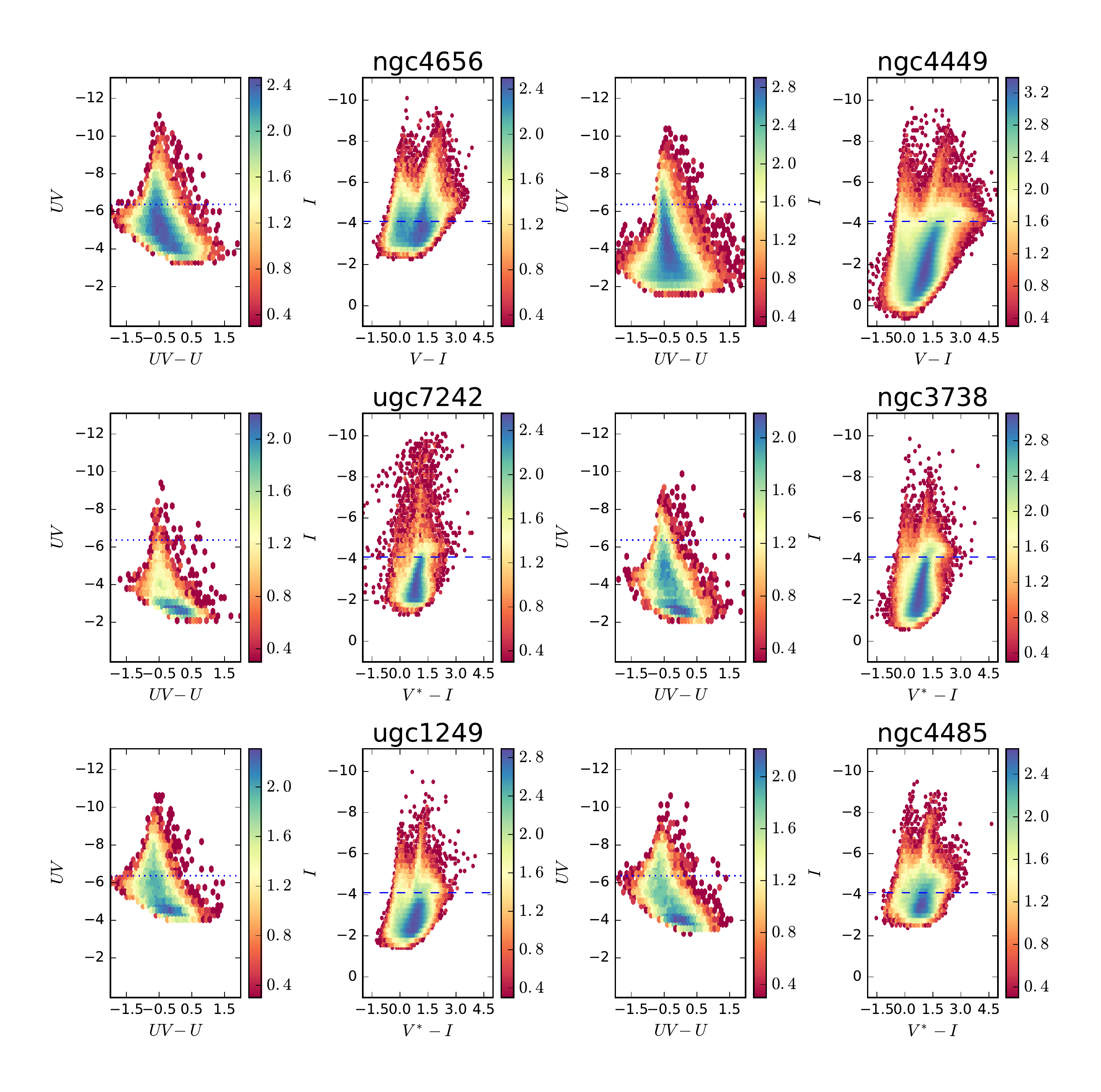}
\caption{\label{f:cmd5} Same as Figure~\ref{f:cmd1}, but for the galaxies NGC4656, NGC4449, UGC7242, NGC3738, UGC1249, NGC4485. Magnitudes marked as V$^*$ refer
to the filter F606W, instead of F5555W.}
\end{figure}

\begin{figure}
\epsscale{1.0}
\plotone{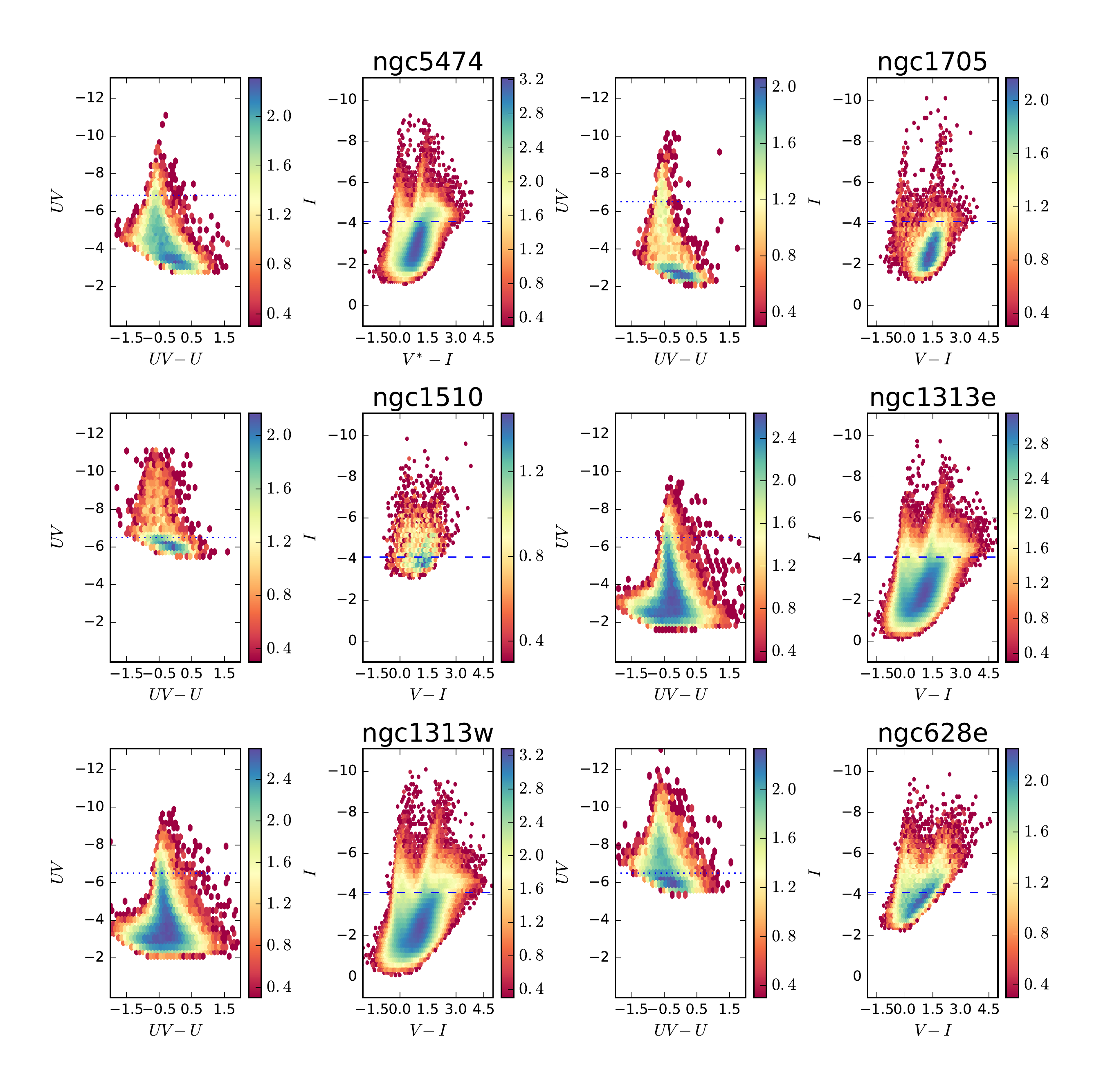}
\caption{\label{f:cmd6} Same as Figure~\ref{f:cmd1}, but for the galaxies NGC5474, NGC1705, NGC1510, NGC1313 (East and West) and NGC628 East. Magnitudes marked as V$^*$ refer
to the filter F606W, instead of F5555W.}
\end{figure}

\begin{figure}
\epsscale{1.0}
\plotone{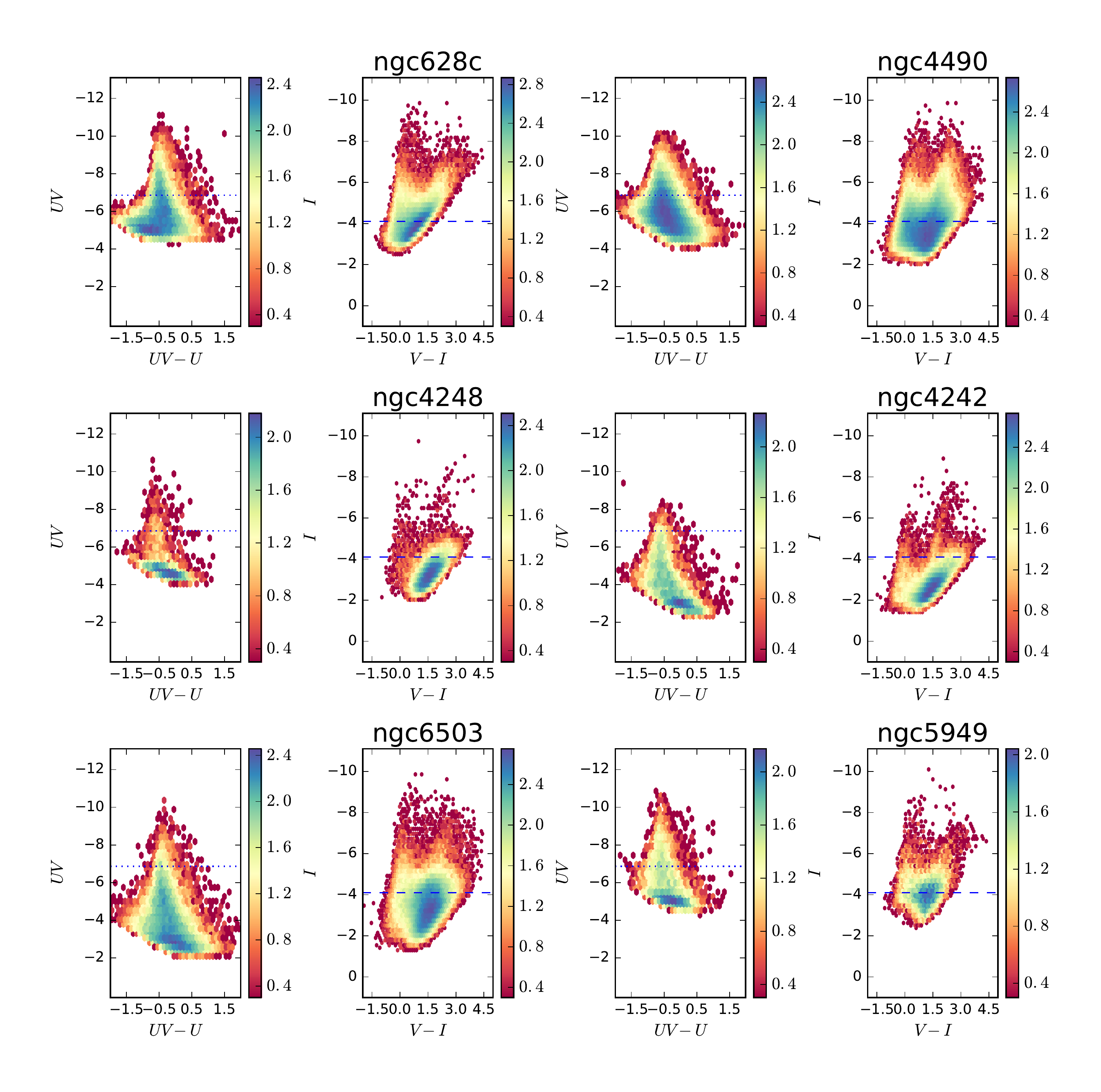}
\caption{\label{f:cmd7} Same as Figure~\ref{f:cmd1}, but for the Western part of the galaxies NGC628 West, NGC4490, NGC4248, NGC4242, NGC6503, NGC5949.}
\end{figure}

\begin{figure}
\epsscale{1.0}
\plotone{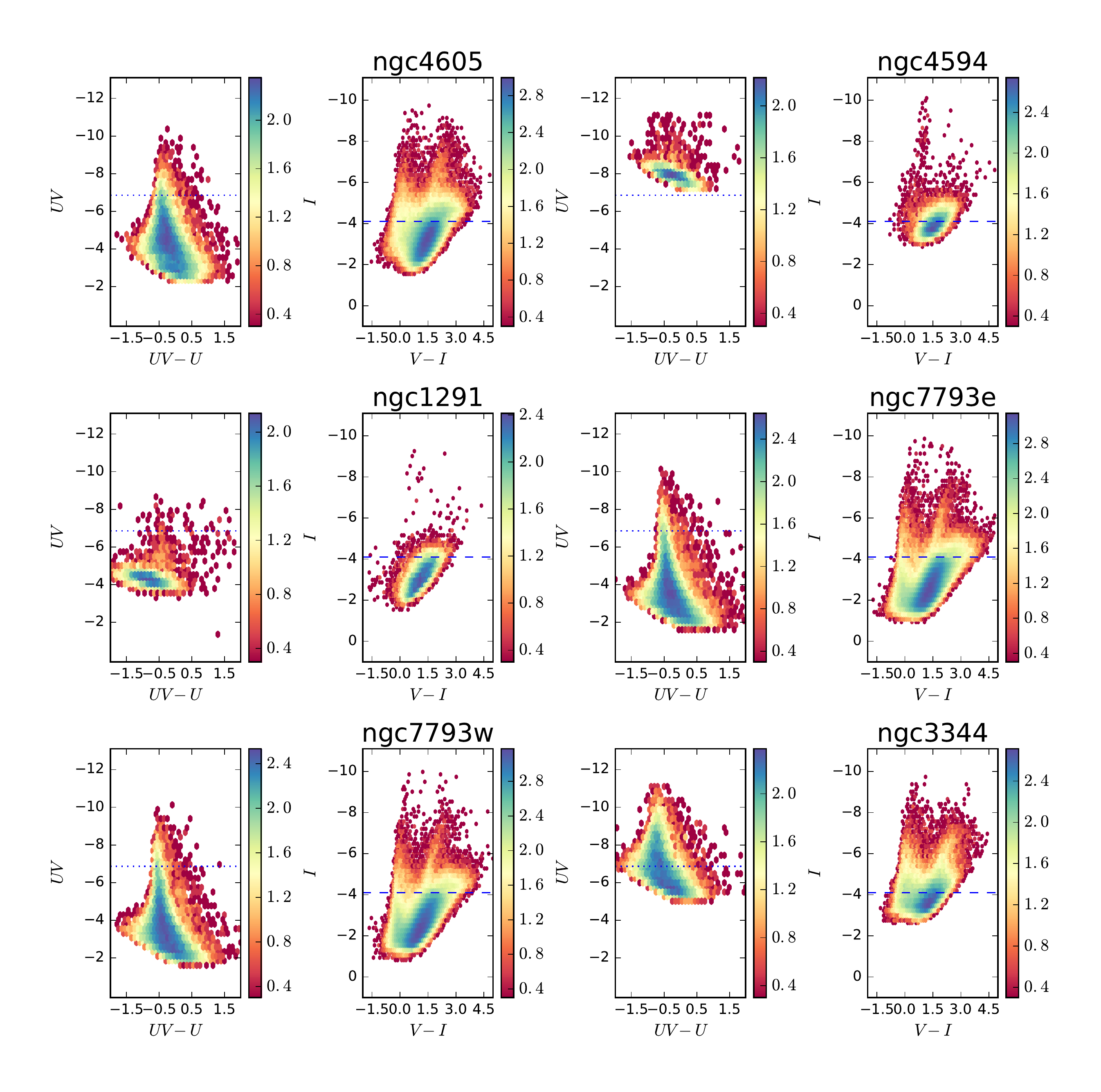}
\caption{\label{f:cmd8} Same as Figure~\ref{f:cmd1}, but for the galaxies NGC4605, NGC4594, NGC1291, NGC7793 (East and  West), and NGC3344.}
\end{figure}

\begin{figure}
\epsscale{1.0}
\plotone{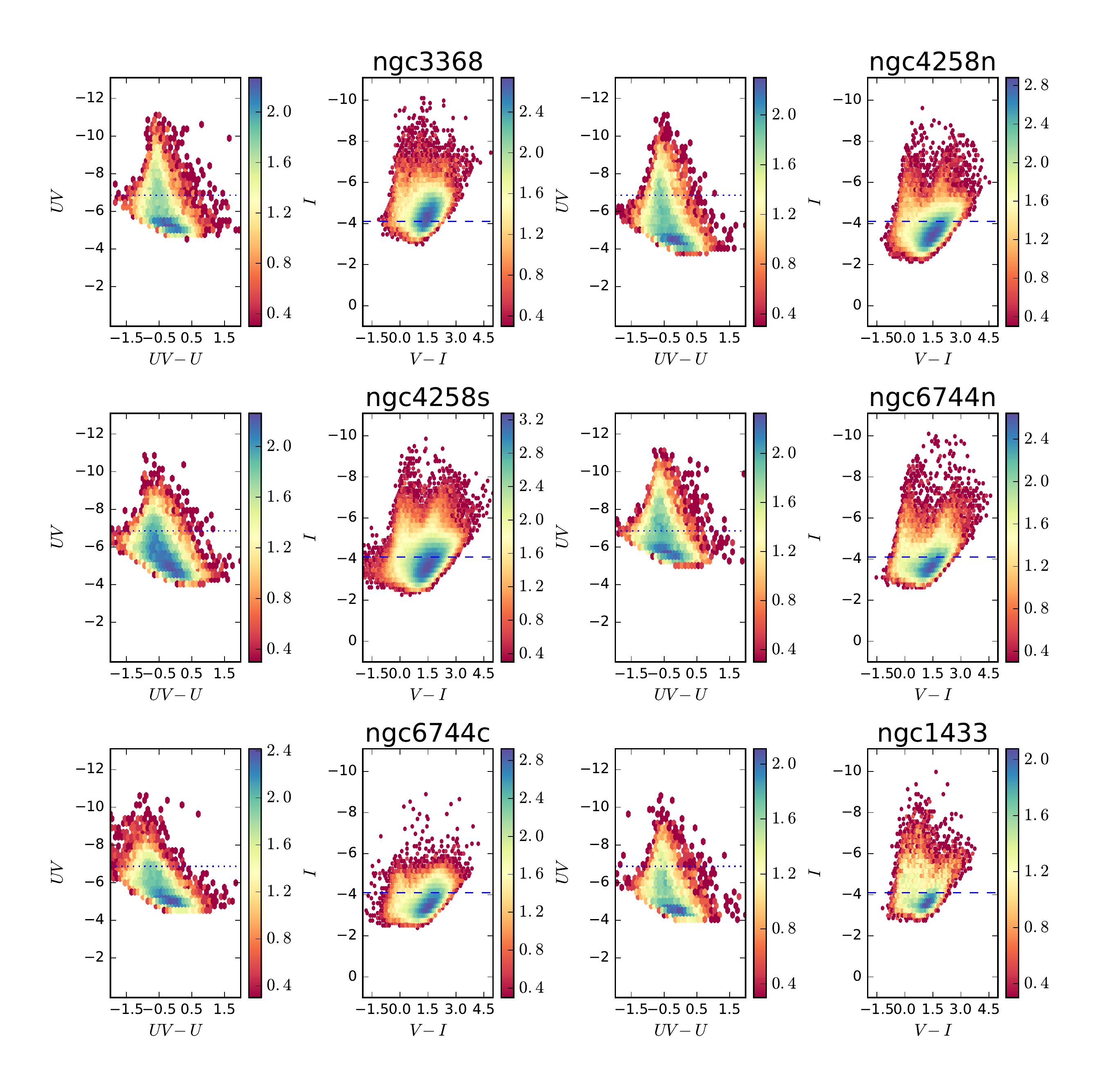}
\caption{\label{f:cmd9} Same as Figure~\ref{f:cmd1}, but for the galaxies NGC3368, NGC4258 (North and South), NGC6744 (North and Center) and NFC1433.}
\end{figure}

\begin{figure}
\epsscale{1.0}
\plotone{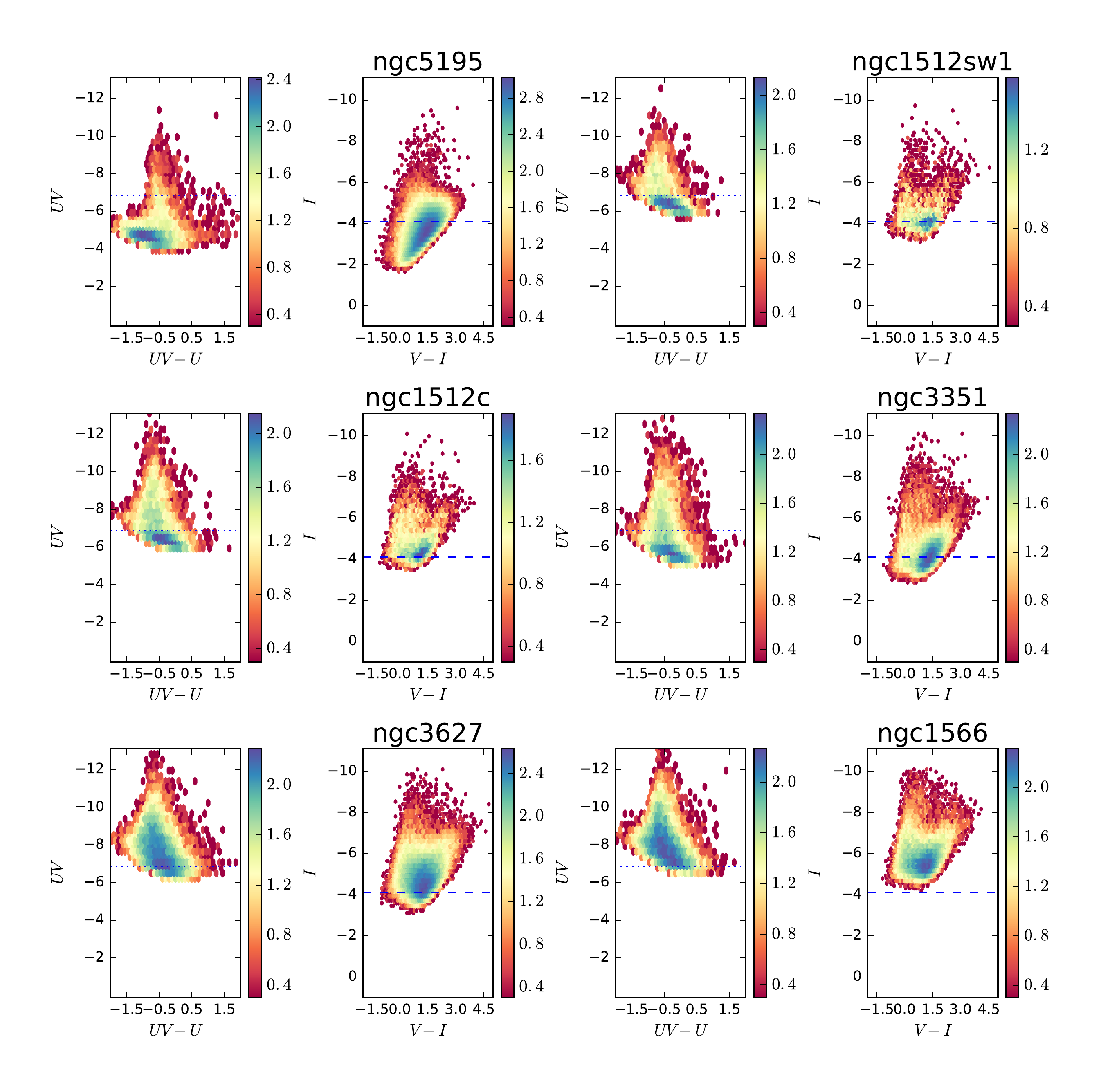}
\caption{\label{f:cmd10} Same as Figure~\ref{f:cmd1}, but for the galaxies NGC5195, NGC1512 (South West and Center), NGC3351, NGC3627, and NGC1566.}
\end{figure}

\begin{figure}
\epsscale{1.0}
\plotone{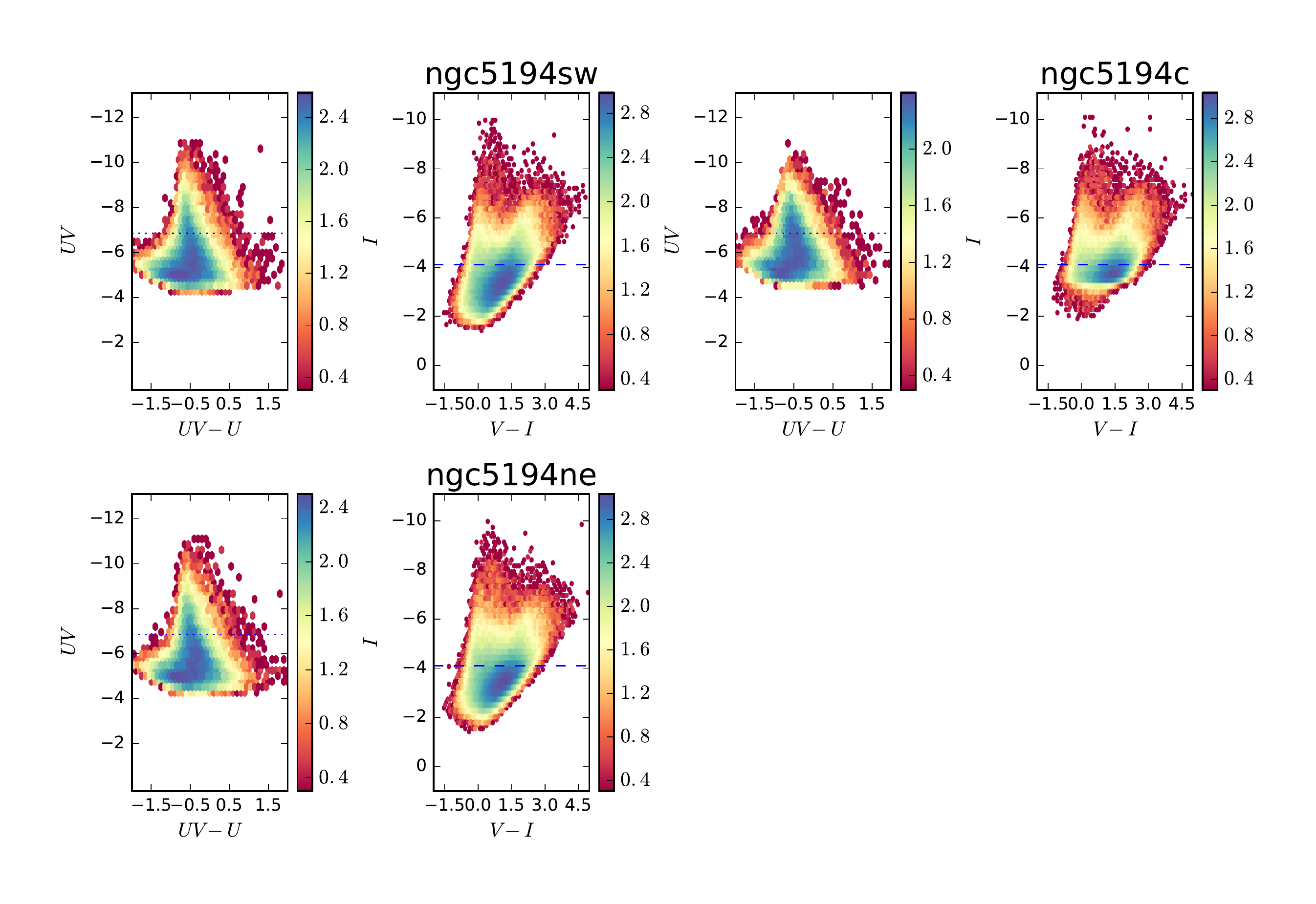}
\caption{\label{f:cmd11} Same as Figure~\ref{f:cmd1}, but for the galaxy NGC5194.}
\end{figure}

\begin{deluxetable}{ccccccccc}
\tablecolumns{9}
\tabletypesize{\scriptsize}
\tablecaption{Log of the Observations\label{log}} 
\tablewidth{0pt}
\tablehead{
\colhead{Data Set } & \colhead{Proposal ID} & \colhead{Obs. Date} & \colhead{Start Time} & \colhead{R.A.} & \colhead{Decl.} & \colhead{Instrument } & \colhead{Filter} & \colhead{Exp. Time}
\\
\colhead{}  & \colhead{} & \colhead{(dd/mm/yy)} & \colhead{(h:m:s)} & \colhead{(2000.0)} & \colhead{(2000.0)} & \colhead{ } & \colhead{} & \colhead{(s)}         }
\startdata
\multicolumn{9}{}{} \\
\multicolumn{4}{}{}& \multicolumn{1}{c}{ESO~486-G021} & \multicolumn{4}{}{} \\
\multicolumn{9}{}{} \\
\hline
 icdm06cnq & 13364 & 06/09/13 & 22:14:43 & $05^d03^m19\fs6$ & $-25\degr25\arcmin22\farcs6$ & WFC3 & F275W & 787 \\
 icdm06cpq & 13364 & 06/09/13 & 22:30:05 & $05^d03^m19\fs7$ & $-25\degr25\arcmin24\farcs4$ & WFC3 & F275W & 787 \\
 icdm06ctq & 13364 & 06/09/13 & 22:45:27 & $05^d03^m19\fs4$ & $-25\degr25\arcmin23\farcs3$ & WFC3 & F275W & 787 \\
 icdm06cyq & 13364 & 06/09/13 & 23:53:39 & $05^d03^m19\fs6$ & $-25\degr25\arcmin22\farcs6$ & WFC3 & F336W & 370 \\
 icdm06d3q & 13364 & 07/09/13 & 00:02:04 & $05^d03^m19\fs7$ & $-25\degr25\arcmin24\farcs4$ & WFC3 & F336W & 370 \\
\enddata
\end{deluxetable}

\end{document}